\documentclass[letterpaper,11pt]{article}

\usepackage[english]{babel}
\usepackage{amssymb}
\usepackage{amsmath}
\usepackage[dvips]{graphicx}
\usepackage{amsthm}
\usepackage{psfrag}
\usepackage[T1]{fontenc}
\usepackage{ae,aecompl}
\usepackage[colorlinks]{hyperref}
\usepackage{subfigure}
\usepackage{appendix}
\usepackage{natbib}
\hypersetup{colorlinks,breaklinks=true,citecolor=blue,filecolor=blue,linkcolor=blue,urlcolor=blue,pdfauthor={C.~Guervilly}}

\newcommand{\BT}{B_{\rm T}}
\newcommand{\BP}{B_{\rm P}}
\newcommand{\bt}{{b_{\rm t}}}
\newcommand{\bp}{{b_{\rm p}}}
\newcommand{\jt}{{j_{\rm t}}}

\newcommand{\ut}{{u_{\rm t}}}
\newcommand{\up}{{u_{\rm p}}}
\newcommand{\JT}{J_{\rm T}}
\newcommand{\JP}{J_{\rm P}}
\newcommand{\cc}{{\rm c}}
\newcommand{\dd}{{\rm d}}
\newcommand{\ee}{{\rm e}}
\newcommand{\ff}{{\rm f}}
\newcommand{\ii}{{\rm i}}
\newcommand{\oo}{{\rm o}}
\newcommand{\rr}{{\rm r}}
\renewcommand{\ss}{{\rm s}}
\newcommand{\ww}{{\rm w}}

\newcommand{\vn}{\boldsymbol{\nabla}}
\newcommand{\vect}[1]{\mathbf{#1}}
\newcommand{\B}{\mathbf{B}}
\newcommand{\vel}{\mathbf{u}}

\newcommand{\pdt}[1]{\frac{\partial #1}{\partial t}}

\newcommand{\pleft}{\left(}
\newcommand{\pright}{\right)}
\newcommand{\Rey}{\textrm{Re}}
\newcommand{\Rm}{\textrm{Rm}}
\newcommand{\Pm}{\textrm{Pm}}

\title{Effect of metallic walls on dynamos generated by laminar boundary-driven flow in a spherical domain}
\author{C\'eline Guervilly$^{1,2}$, Toby S. Wood$^{1,2}$ \& Nicholas H. Brummell$^2$ \vspace{0.2cm}
 \\ {\small $ ^1$Department of Applied Mathematics,}
 \\ {\small University of Leeds, Leeds LS2 9JT, UK}
 \\ {\small $ ^2$Department of Applied Mathematics and Statistics,}
 \\ {\small  University of California, Santa Cruz, CA95064, USA}}

\begin{document}

\maketitle

\begin{abstract}
We present a numerical study of dynamo action in a conducting fluid encased in a metallic spherical shell.
Motions in the fluid are driven 
by differential rotation of the outer metallic shell, 
which we refer to as ``the wall''.
The two hemispheres of the wall are held in counter-rotation,
producing a steady, axisymmetric interior flow consisting of
differential rotation and a two-cell meridional circulation with 
radial inflow in the equatorial plane.
From previous studies, this type of flow is known to maintain a stationary equatorial dipole by dynamo action  
if the magnetic Reynolds number is larger than about 300 and if
the outer boundary is electrically insulating.
We vary independently the thickness, electrical conductivity,
and magnetic permeability of the wall to determine their effect on the dynamo action. 
The main results are:
(a) Increasing the conductivity of the wall hinders the dynamo 
by allowing eddy currents within the wall, which are induced by 
the relative motion of the equatorial dipole field and the wall.
This processes can be viewed as a skin effect or, equivalently,
as the tearing apart of the dipole by the differential rotation
of the wall, to which the field lines are anchored by
high conductivity.
(b) Increasing the magnetic permeability of the wall favors dynamo action
by constraining the magnetic field lines in the fluid to be normal to the wall,
thereby decoupling the fluid from any
induction in the wall.
(c) Decreasing the wall thickness limits the amplitude of the eddy currents,
and is therefore favorable for dynamo action,
provided that the wall is thinner than the skin depth.
We explicitly demonstrate these effects of the wall properties on the dynamo field
by deriving an effective boundary condition in the
limit of vanishing wall thickness.
\end{abstract}

\section{Introduction}

Many planets and stars have observable surface magnetic fields that
are generated by
hydromagnetic dynamo action in their deep interior, where the motions of a conducting fluid, 
either ionized plasma or liquid metal,
maintain the magnetic field against ohmic diffusion.
In such systems, the conducting fluid is typically surrounded by an electrically insulating medium
that plays no role in the dynamo process.
In the last decade, a number of laboratory experiments have been constructed to study dynamo 
action in either turbulent or laminar flow regimes. 
Most of these experiments use liquid sodium and drive flows by mechanical forcing at the boundaries.
So far, three experiments have successfully generated magnetic field by fluid motions 
\citep{Gai01,Sti01,Mon07}. 
Of these successful experiments, the von K\'arm\'an Sodium (VKS) experiment offers the closest 
approximation to a natural dynamo, in the sense that 
a large scale magnetic field is generated by
a relatively unconstrained, highly turbulent flow
driven by the counter-rotation of two impellers.
However, at the experimental parameters attainable, dynamo action is only
observed when the impellers are made of soft iron, and not when they are made of stainless steel.  
Iron has a higher magnetic permeability than steel,
and so these results imply that 
magnetic boundary conditions play a crucial role in
the VKS experiment, and possibly in other problems involving magnetohydrodynamics (MHD).
Understanding the effect of magnetic boundary conditions is therefore essential to interpreting 
the results of upcoming dynamo experiments, such as
the plasma experiment in Madison,
Wisconsin \citep{Spe09}, and the spherical-Couette liquid sodium experiment in College Park, 
Maryland \citep{Zim11}.

Motivated by the above, in this paper we use numerical simulations to
investigate the effect of magnetic boundary conditions on dynamo action 
produced by an axisymmetric \emph{laminar} flow driven by an azimuthal boundary forcing in spherical geometry.
This geometry is particularly relevant to the Madison plasma experiment. 
Here, we consider only the case where the azimuthal boundary forcing is anti-symmetric with respect to the equatorial plane,
in which case an axisymmetric poloidal circulation is also established with a single meridional cell in each hemisphere.
The ability of a flow with these symmetry properties to maintain
a magnetic field was first established by \citet{Gub73},
who found that the critical magnetic Reynolds number
(the minimum ratio of the magnetic diffusion and induction timescales
required for dynamo action) was about $50$.
Subsequent studies
\citep[e.g.][]{Dud89,Nak91,Holme03,Mar03}
confirmed that the most readily generated magnetic field for this type of flow
is an equatorial dipole.
These studies were all kinematic, in the sense that the velocity field
was prescribed, rather than determined dynamically and self-consistently.
This allowed the structure of the flow, including the ratio
of its poloidal and toroidal components, to be varied arbitrarily.
In a more realistic model for which the flow is driven by viscous
drag at a rotating outer boundary, the ratio of poloidal to toroidal flow is not an
adjustable parameter, but rather is determined by the dynamics of the system,
and is a function of the Reynolds number 
(the ratio of the viscous diffusion timescale to the boundary forcing timescale).
Self-consistent numerical simulations are therefore required to determine the dynamo
properties of these laminar shear flows.

Using fully dynamical (i.e.~non-kinematic) spherical simulations and an azimuthal boundary 
forcing thought to be achievable in the Madison plasma experiment,
\citet{Spe09} obtained a dynamo for a magnetic Reynolds number 
of about $300$ when using electrically insulating boundary conditions,
which assume that electric currents vanish everywhere outside the fluid.
However, simulations run with the same boundary forcing and magnetic Reynolds number failed to produce a dynamo
when perfectly electrically conducting boundary conditions were used
(Forest, private communication).
This result was somewhat unexpected, because in numerical simulations run with different
types of shear flows, notably in turbulent regimes, increasing the conductivity
of the outer wall is usually favorable to dynamo action 
\citep[e.g.][]{Kai99,Ava03,Lag08,Rob10,Gue12}.  
In an attempt to bridge the gap between the two
idealized limits of true physical boundaries
used in \citeauthor{Spe09},
\citet{Kha12} used
the same flow, but applied boundary conditions derived in the ``thin-wall limit''
in which the outer wall thickness $h$ tends to zero, but either the
integrated conductivity $h \sigma$ or the integrated permeability $h \mu$
remains finite (see \citet{Rob10} and detailed discussion later in this paper).
They found that varying $h \sigma$ has no effect on the value of the critical 
magnetic Reynolds number for dynamo action, contrary to what might have been expected from the results of
\citeauthor{Spe09}. However, as $h \sigma$ is increased the growth rate of the magnetic field tends to zero.
Increasing  $h \mu$ has a positive effect on the dynamo
action as the critical magnetic Reynolds decreases
by about 35\% for $h \mu \to \infty$ compared to its value for $h \mu \to 0$. 

The purpose of the present paper is to unify and interpret these previous results, 
and to reach a full understanding of the role of the wall in fully dynamical dynamo simulations using the same
laminar flow as in \citet{Spe09}.
Rather than employing an approximation for the effect of the metallic wall
on the magnetic field,
we here include a wall of finite thickness in the computational domain, which allows us to vary independently 
the wall thickness, electrical conductivity, and magnetic permeability.

After outlining our numerical model in Section~\ref{sec:num},
we present the results 
of simulations performed for various values of the wall thickness,
electrical conductivity, and magnetic permeability.
We provide a physical interpretation for the effect of the wall properties
on the dynamo mechanism.
In Section~\ref{sec:analytical_sol}, we describe a new magnetic boundary 
condition derived in the thin-wall limit,
which generalizes those of \citet{Rob10} and \citet{Kha12}, and which
further elucidates the effects of the wall on the dynamo.

\section{Numerical model}
\label{sec:num}
\begin{figure}
  \centering
   \subfigure[]{\label{fig:model3d}
   \includegraphics[clip=true,width=8cm]{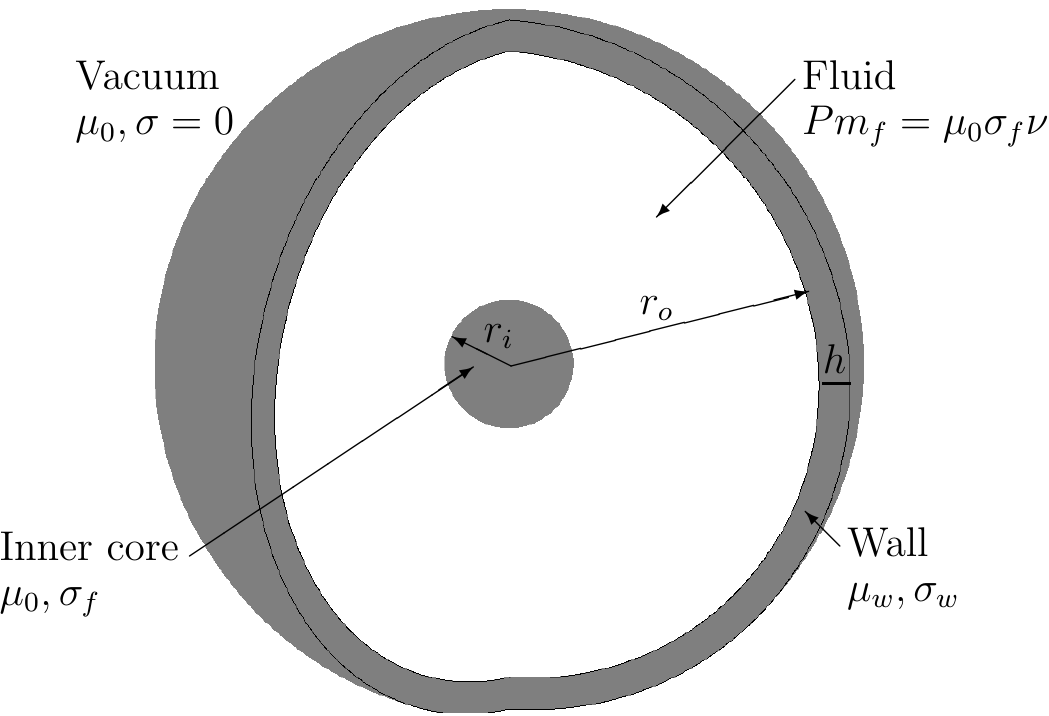}}
  \subfigure[]{\label{fig:uphi_BC}
  \includegraphics[clip=true,width=6cm]{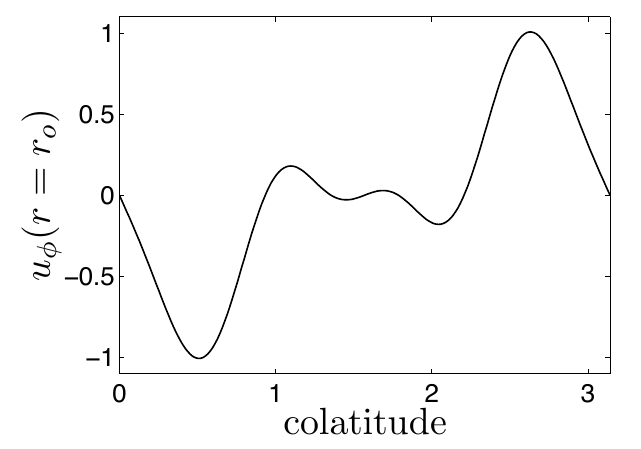}}
  \caption{(a) 3D view of the model. The gray areas represent the inner core and the outer wall. 
(b) Azimuthal velocity imposed at $r=r_\oo$ with $U_\ww=1$.}
\end{figure}

We use the same three-dimensional, fully non-linear numerical code
that was described in detail by \citet{Gue12}. 
Here, we describe only the mathematical details of the model, 
and refer the reader to \citeauthor{Gue12} 
for more details on the numerical algorithm.
Figure~\ref{fig:model3d} presents a 3D schematic view of the model.
The domain is spherical and consequently we express all fields in
spherical coordinates $(r, \theta, \phi)$, with $r$ the radius, $\theta$ the colatitude, 
and $\phi$ the azimuthal angle or longitude.
An electrically conducting, incompressible fluid fills the spherical shell between an inner radius
$r_\ii$ and an outer radius $r_\oo$. The fluid has viscosity $\nu$,
density $\rho$, electrical conductivity $\sigma_\ff$, and magnetic permeability equal
to that of the vacuum, $\mu_0$. 
The fluid properties are assumed to be uniform and are kept fixed throughout this study.
The fluid is surrounded by an outer spherical shell or ``wall'' of finite thickness $h$, 
and has uniform electrical conductivity $\sigma_\ww$ and
uniform magnetic permeability $\mu_\ww$. 

An angular velocity profile varying with latitude, $\Omega(\theta)$, is prescribed in the wall. 
We impose impenetrable and no-slip boundary conditions at the inner edge of the wall, $r=r_\oo$, 
so the viscous stress exerted by the wall drives an axisymmetric azimuthal flow in the fluid.
The radial and latitudinal components of the velocity are set to zero in the wall.
Since the wall has a fixed shape, and is impenetrable to the fluid, it is convenient to think of it as solid
even though it is not in solid body rotation.
Our differentially rotating wall
allows us to approximate the boundary driving in various laboratory experiments,
such as in the upcoming Madison plasma experiment. 
We note that our numerical model differs somewhat from that of \citet{Kha12}.
In particular, in their model, the boundary condition for the magnetic field assumes that 
the outer wall is at rest in the laboratory frame. 
For reasons discussed in Section~\ref{sec:analytical_sol},
the additional complexity of the wall properties in the model of \citet{Kha12}
leads to some differences between their results and ours.

For the azimuthal velocity of the wall, 
we choose the same latitudinal profile as \citet{Spe09},
\begin{eqnarray}
   u_{\phi}(r,\theta) = U_\ww \frac{r}{r_\oo} \sum\limits_k C_k \sin (k \theta)
   \hspace{1cm}\mbox{for}\hspace{1cm}r_\oo<r<r_\oo+h,
\label{eq:forcing}
\end{eqnarray}
where the constant $U_\ww$ is a characteristic forcing velocity and
\begin{equation}
	C_2 = -0.4853, \quad C_4 = -0.5235, \quad C_6=-0.0467, \quad \mbox{and} \quad C_8=0.1516.
\end{equation}
The other coefficients $C_k$ in Eq.~(\ref{eq:forcing}) are all set to zero.
The azimuthal velocity at $r=r_\oo$ is plotted in Fig.~\ref{fig:uphi_BC} for $U_\ww=1$. 
The azimuthal velocity is anti-symmetric about the equator and is mostly localized
at mid-latitudes.

Within the fluid, we solve the incompressible MHD equations:
\begin{eqnarray}
	&& \pdt{\vect{u}} + \vect{u} \cdot \nabla \vect{u} = -\frac{1}{\rho} \nabla p + \nu \nabla^2 \vect{u} + \frac{1}{\rho} \vect{J} \times \vect{B},
	\\
	&& \nabla \cdot \vect{u} = 0, 
	\label{eq:divu}
	\\
	&&\pdt{\vect{B}} = \nabla \times \pleft \vect{u} \times \vect{B} \pright - \nabla \times \frac{1}{\sigma} \nabla \times \frac{\vect{B}}{\mu},
	\label{eq:induction}
	\\
	&&\nabla \cdot \vect{B} = 0,
	\label{eq:divb}
\end{eqnarray}
where $\vect{u}$ is the velocity, $p$ is the pressure,
$\vect{B}$ is the magnetic field, 
\mbox{$\vect{J}=\nabla \times (\vect{B}/\mu)$} is the electric current density, and with
$\sigma = \sigma_\ff$ and $\mu=\mu_0$ in the fluid.

Within the wall, we solve only the magnetic induction equation (Eq.~(\ref{eq:induction})) 
using the prescribed velocity in the wall (Eq.~(\ref{eq:forcing})) and with
$\sigma=\sigma_\ww$ and $\mu=\mu_\ww$. 

The region outside the wall, $r>r_\oo+h$, is assumed to be a perfect vacuum, for which 
$\sigma=0$ and $\mu=\mu_0$.
In this region, the field is determined analytically by solving \mbox{$\nabla \times \B/\mu_0 = 0$},
so that, numerically, the vacuum is treated as a boundary condition at $r=r_\oo+h$.

For numerical convenience, we impose a solid inner core at the center for $r\leq r_\ii$ with \mbox{$r_\ii=0.05r_\oo$}.
The inner core is held at rest and has the same electrical conductivity and magnetic permeability as the fluid. 
We solve the magnetic induction equation within the inner core 
with zero velocity. The boundary conditions for the velocity at \mbox{$r=r_\ii$} are no-slip and impenetrable.

The equations are solved in non-dimensional form. 
The length is scaled by the outer radius of the fluid $r_\oo$, the velocity by the forcing velocity amplitude $U_\ww$,
the time by $r_\oo/U_\ww$, and the magnetic field by \mbox{$U_\ww(\rho \mu_0)^{1/2}$}.
The dimensionless parameters for the fluid are the magnetic Prandtl number:
\begin{eqnarray}
	\Pm_\ff=\mu_0 \sigma_\ff \nu,
\end{eqnarray}
and the Reynolds number, which is the ratio of the viscous timescale $r_\oo^2/\nu$ to the forcing
timescale $r_\oo/U_\ww$:
\begin{eqnarray}
	\Rey=\frac{U_{w} r_\oo}{\nu}.
\end{eqnarray}
All of the simulations presented here have $\Pm_\ff=1$ and $\Rey=300$.
The dimensionless parameters for the wall are the relative wall thickness \mbox{$\hat{h}=h/r_\oo$}, the relative conductivity
\mbox{$\sigma_\rr=\sigma_\ww/\sigma_\ff$}, and the relative magnetic permeability \mbox{$\mu_\rr=\mu_\ww/\mu_0$}.

At the fluid--wall interface, and at the wall--vacuum interface,
the electrical conductivity and magnetic permeability
are discontinuous, leading to the following matching conditions for
the normal and tangential components of
$\vect{B}$ and $\vect{J}$:
\begin{eqnarray}
	&& \vect{B}^+ \cdot \vect{e}_r = \vect{B}^- \cdot \vect{e}_r,
	\label{eq:Bnormal}
	\\
	&& (\vect{B}/\mu)^+ \times \vect{e}_r = (\vect{B}/\mu)^- \times \vect{e}_r,
	\label{eq:Btan}
	\\
	&& \vect{J}^+ \cdot \vect{e}_r = \vect{J}^- \cdot \vect{e}_r,
	\label{eq:jnormal}
	\\
	&& (\vect{J}/\sigma)^+ \times \vect{e}_r = (\vect{J}/\sigma)^- \times \vect{e}_r,
	\label{eq:jtan}
\end{eqnarray}
where the superscripts $-$ and $+$ indicate, respectively, values immediately inside and outside the interface.

We use a poloidal--toroidal representation
for the velocity and magnetic fields
in order to enforce the divergence-free conditions
(Eqs.~(\ref{eq:divu}) and~(\ref{eq:divb})).
For the magnetic field, we define poloidal and toroidal scalar potentials $\BP$ and $\BT$ such that
\begin{equation}
	\B = \nabla \times \nabla \times (\BP \vect{r} ) + \nabla \times (\BT \vect{r} ).
    \label{eq:Bpoltor}
\end{equation}
The spherical components of the magnetic field are then 
\begin{eqnarray}
	B_r &=& \frac{1}{r} L_2 \BP, \label{eq:Br_Bp}
	\\
	B_{\theta} &=& \frac{\partial}{\partial \theta} \frac{1}{r} \frac{\partial}{\partial r} r \BP
                     + \frac{1}{\sin \theta} \frac{\partial \BT}{\partial \phi},
          \label{eq:Btheta}
	\\
	B_{\phi} & = & \frac{1}{\sin \theta} \frac{\partial}{\partial \phi} \frac{1}{r} \frac{\partial}{\partial r} r \BP
                     - \frac{\partial \BT}{\partial \theta},
          \label{eq:Bphi}
\end{eqnarray}
where the angular laplacian operator $L_2$ is defined as
\begin{equation}
	L_2 = - \frac{1}{\sin \theta} \frac{\partial}{\partial \theta} \pleft \sin\theta \frac{\partial}{\partial \theta} \pright 
	- \frac{1}{\sin^2\theta}\frac{\partial^2}{\partial \phi^2}.
  \label{eq:L2}
\end{equation}
Note that the toroidal magnetic field has no radial component.

The boundary conditions (Eqs.~(\ref{eq:Bnormal}) and~(\ref{eq:Btan})) 
at $r=r_\oo$ imply that 
\begin{eqnarray}
	\BP(r_\oo^+) &=& \BP (r_\oo^-), \label{eq:cont_bp}
	\\
	\BT(r_\oo^+) & = & \mu_\rr \BT(r_\oo^-), \label{eq:cont_bt}
	\\
	\left. \frac{\partial r \BP}{\partial r} \right|_{r_\oo^+} & = & 
	\mu_\rr \left. \frac{\partial r \BP}{\partial r} \right|_{r_\oo^-}. \label{eq:cont_bs}
\end{eqnarray}

It is sometimes convenient to also represent
the electric current density $\vect{J}$ in terms of poloidal and toroidal scalar potentials 
$\JP$ and $\JT$, which are related to $\BT$ and $\BP$ by
\begin{eqnarray}
	\JP &=& \frac{\BT}{\mu}, \label{eq:def_jp}
	\\
	\JT &=&
    \frac{1}{r^2} L_2 \frac{\BP}{\mu}
    - \frac{1}{r}\frac{\partial}{\partial r}\frac{1}{\mu}\frac{\partial r \BP}{\partial r}.
    \label{eq:def_jt}
\end{eqnarray}
The boundary conditions~(\ref{eq:jnormal}) and~(\ref{eq:jtan}) then become 
\begin{eqnarray}
	\JP(r_\oo^+) &=& \JP (r_\oo^-), \label{eq:cont_jp}
	\\
	\JT(r_\oo^+) & = & \sigma_\rr \JT(r_\oo^-), \label{eq:cont_jt}
	\\
	\left. \frac{\partial r \JP}{\partial r} \right|_{r_\oo^+} 
	& = & \sigma_\rr \left. \frac{\partial r \JP}{\partial r} \right|_{r_\oo^-}.  \label{eq:cont_js}
\end{eqnarray}
We note that the poloidal (toroidal) field is produced
by the toroidal (poloidal) current, and that
the toroidal electric current $\nabla \times (\JT \vect{r})$ has no radial component.

We expand the poloidal and toroidal scalars in spherical harmonics $Y_l^m$ in the angular coordinates,
where $l$ represents the latitudinal degree and $m$ the azimuthal order:
\begin{eqnarray}
	\BP(r,\theta,\phi,t) &=& \sum_{l=0}^{l_{\rm max}} \sum_{m=0}^{m_{\rm max}}  \bp_l^m(r,t)  Y_l^m(\theta, \phi),
	\\
	\BT(r,\theta,\phi,t) &=& \sum_{l=0}^{l_{\rm max}} \sum_{m=0}^{m_{\rm max}}  \bt_l^m(r,t) Y_l^m(\theta, \phi),
\end{eqnarray}
and similarly for the poloidal and toroidal scalars of the velocity.

Our numerical code evolves the quantities $\bp_l^m(r,t)$, etc.~using a second-order 
finite difference scheme on an irregular radial grid.
For further details see \citet{Gue12}.
For the laminar flow simulations presented here, the numerical resolution has been taken as 
$300$ radial points in the fluid, between $20$ and $50$ radial points in the wall depending on the
wall parameters, and $10$ radial points in the inner core.
The spherical harmonic expansion is truncated at
\mbox{$l_{\rm max}=64$} and \mbox{$m_{\rm max}=12$}.
For the laminar flow considered here, the
kinetic and magnetic energy spectra in $l$ and $m$ are well resolved at this resolution,
and a finer radial resolution does not change the numerical solution significantly.
Each simulation is integrated in time until
the kinetic and magnetic energies reach stationary values (see Fig.~\ref{fig:ME_vs_time}).

\section{Results}

\subsection{General characteristics}
\label{sec:dynamo}

\begin{figure}
 \centering
   \includegraphics[clip=true,height=6.cm]{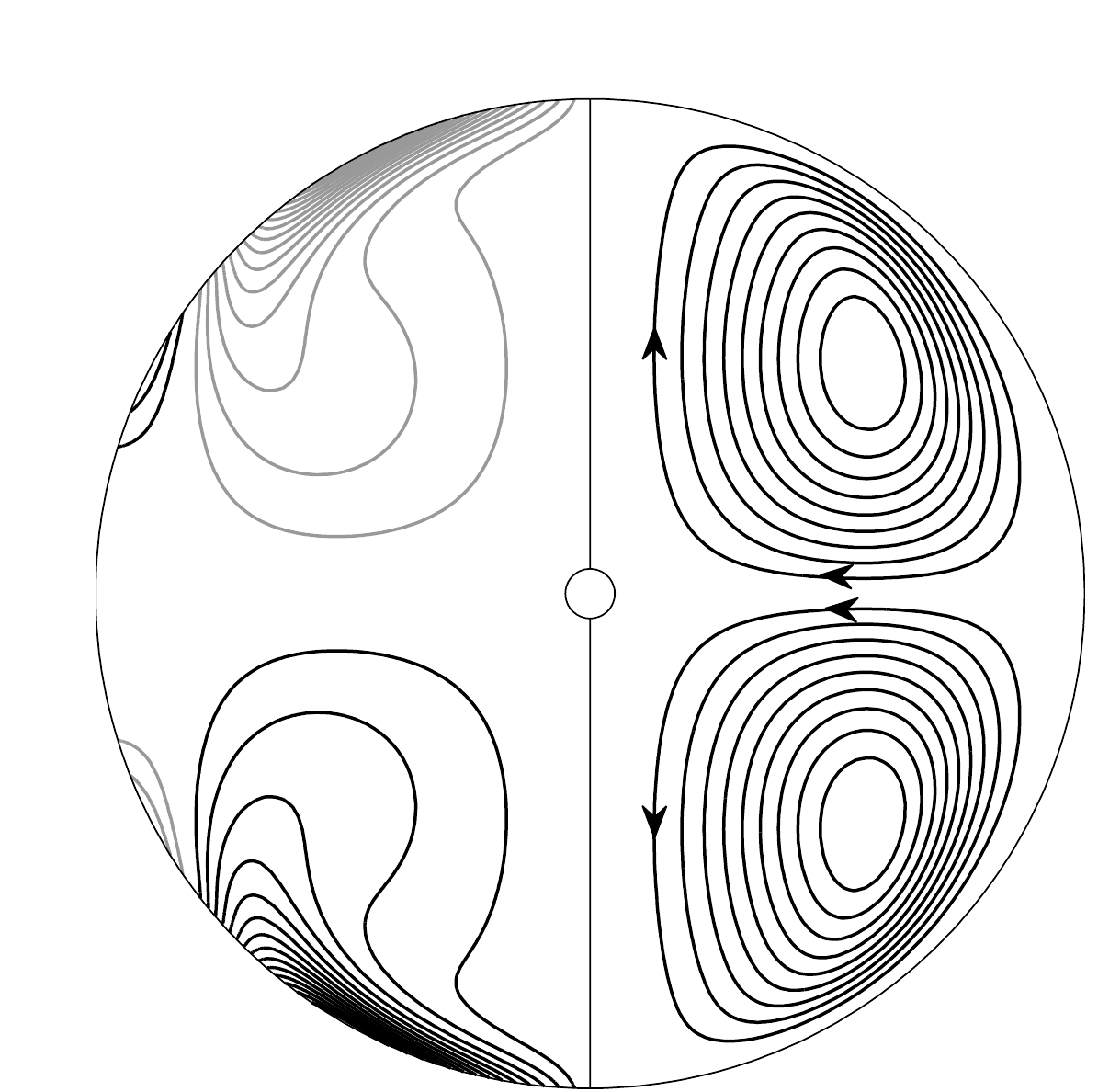}
  \caption{Axisymmetric flow in a meridional plane for $\Rey=300$.
  Left half: azimuthal velocity, where black indicates a positive value, and gray a negative value.
  Right half: poloidal streamlines, with direction of circulation indicated.}
  \label{fig:mean_flow}
\end{figure}

The differential rotation $\Omega(\theta)$ of the wall 
drives an axisymmetric azimuthal velocity in the fluid
through viscous drag at $r=r_\oo$. The differential rotation also establishes
an axisymmetric poloidal circulation consisting of one meridional cell in
each hemisphere with inward radial flow in the equatorial plane.
In the absence of a magnetic field,
this flow is hydrodynamically stable for $\Rey \lesssim 500$
\citep{Spe09}.
All of the simulations presented in this paper have $\Rey=300$,
so we expect the flow to be predominantely steady and axisymmetric
(although the Lorentz force from the magnetic field can drive some
nonaxisymmetric flow).
The flow from a typical simulation is illustrated in Fig.~\ref{fig:mean_flow},
which displays contours of the azimuthal velocity
and streamlines of the poloidal circulations in the meridional plane.

\begin{figure}
 \centering
   \includegraphics[clip=true,width=6cm]{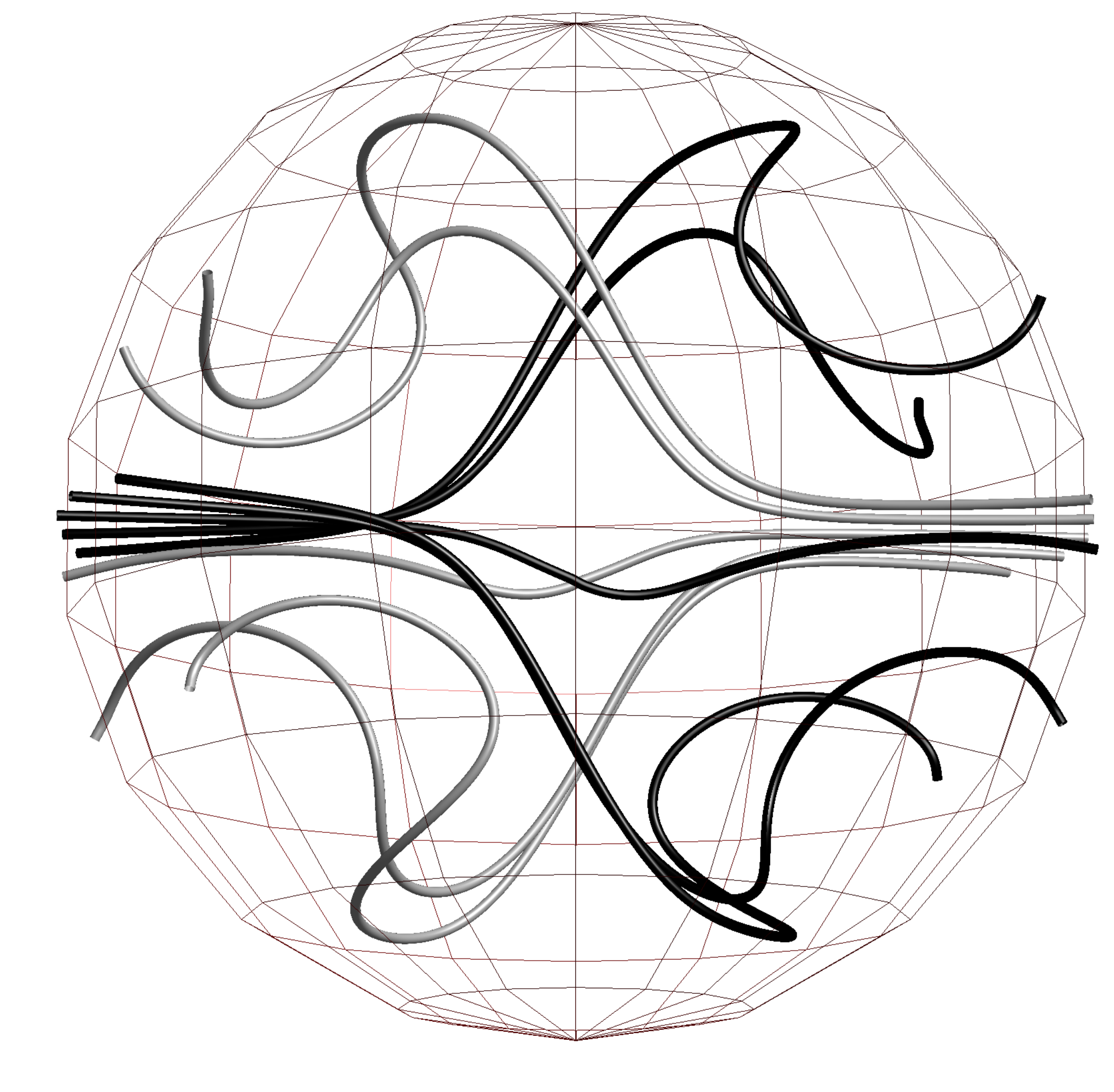}
  \caption{Magnetic field lines in the fluid (black and gray lines) for the dynamo simulation at
  $(\hat{h}, \sigma_\rr, \mu_\rr)=(0.1, 10^{-3},1)$.
  Gray and black lines start from points at $r=r_\oo$ where the radial magnetic field is 
  positive or negative respectively. The mesh shows the 
  outer sphere at $r=r_\oo$. The magnetic field lines are not plotted in the wall or vacuum.}
  \label{fig:B_3D}
\end{figure}

An axisymmetric flow cannot maintain
an axisymmetric magnetic field \citep{Cow33},
but can potentially maintain a \emph{non}axisymmetric magnetic field.
In fact, with $\Rey=300$ and $\Pm_\ff=1$, 
and for certain choices of magnetic boundary conditions, 
the flow maintains a steady magnetic field for which $\BP$ and $\BT$ are both
dominated by the spherical harmonics of degree $l=1$ and order $m=1$.
The $(l,m)=(1,1)$ poloidal field corresponds to an equatorial dipole.
This field configuration is common to all the dynamo cases presented here.
Figure~\ref{fig:B_3D} shows a three-dimensional view of the magnetic field lines
for the dynamo case \mbox{$(\hat{h},\sigma_\rr, \mu_\rr)=(0.1,10^{-3},1)$}, 
which we refer to as ``Case D'' hereafter.
Similar magnetic field configurations have been obtained in previous
numerical simulations using this type of axisymmetric shear flow 
\citep[e.g.][]{Mar03,Bay07,Gis08b}.

Before considering the effect of changing the wall properties in the next sections,
we first discuss the dynamo mechanism.

Using the selection rules described by \citet{Bul54}, 
the action of a laminar flow on the different components of the magnetic field, 
as expressed by Eq.~(\ref{eq:induction}), can be deduced from 
its symmetry properties.
We represent the velocity field in terms of poloidal and toroidal scalars, $\up_l^m$ and $\ut_l^m$.
The north--south symmetry properties of our predominantly
axisymmetric ($m=0$) flow imply
that only the spherical components of the velocity
with even degree $l$ are non-zero.
Moreover, we find in all our simulations that the magnetic field is
dominated by its $m=1$ components.
For simplicity, we therefore consider only 
the action of the components of the velocity with even degree $l$ and $m=0$
on the $m=1$ poloidal and toroidal magnetic fields.
To simplify the notation in the rest of the paper, we omit
the superscript $m$ for the spectral coefficients.

The selection rules partition the magnetic field into two orthogonal families:
$(\bp_\oo^\cc, \bp_\ee^\ss, \bt_\oo^\cc, \bt_\ee^\ss)$
and $(\bp_\oo^\ss, \bp_\ee^\cc, \bt_\oo^\ss, \bt_\ee^\cc)$,
where the subscripts $\oo$ and $\ee$ denote odd and even degrees in $l$ respectively, and  
the superscripts $\cc$ and $\ss$ denote the real and imaginary parts of the spectral coefficients 
(i.e.,~$\cos(\phi)$ and $\sin(\phi)$) respectively.
Which family dominates the solution depends only on the initial conditions for the magnetic field.

\begin{figure}
 \centering
   \subfigure[]{\label{fig:dynamo_mechanism}
   \includegraphics[clip=true,width=5cm]{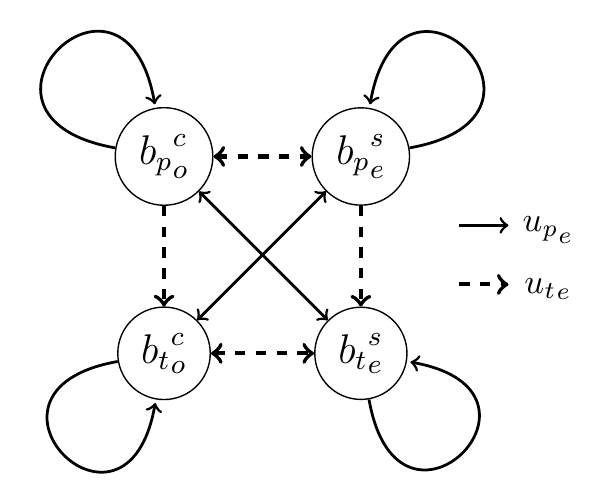}}
   \subfigure[]{\label{fig:dynamo_loops}
   \includegraphics[clip=true,width=7cm]{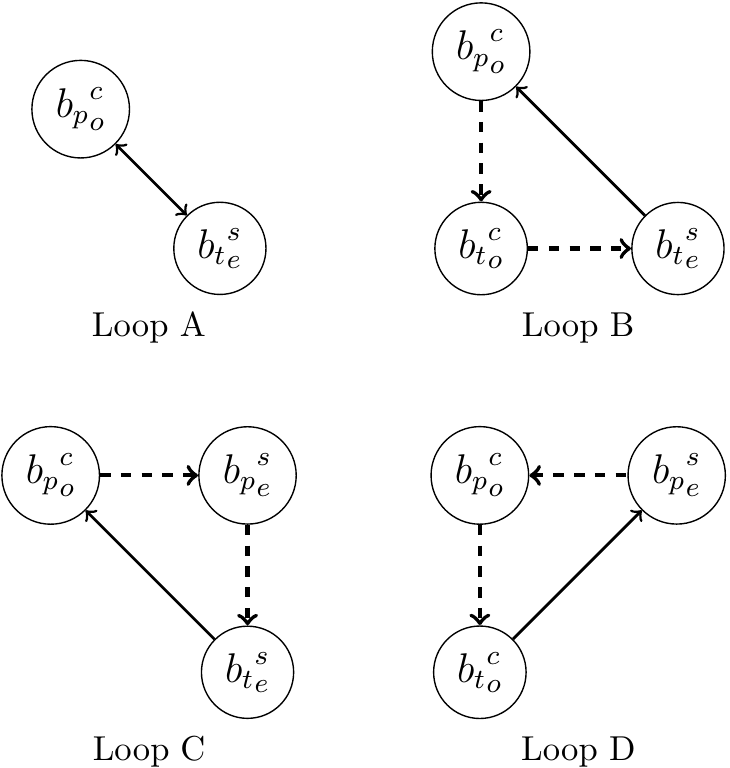}}
  \caption{Bullard \& Gellman diagram \citep{Bul54} for the flow considered here.
  (a) Action of the toroidal flow (dashed arrows) and the poloidal flow (solid arrows)
on the different components of an $m=1$ magnetic field, decomposed into poloidal and toroidal parts
($\bp$ and $\bt$ respectively) and odd and even degrees $l$ (subscripts $\oo$ and $\ee$ respectively).
Straight arrows indicate transfers between different magnetic components, and closed arrows 
indicate rearrangement of a magnetic component.
(b) Possible loops for the dynamo mechanism.}
\end{figure}

Figure~\ref{fig:dynamo_mechanism} shows schematically how
the poloidal and toroidal components of the velocity,
$\up_\ee$ and $\ut_\ee$ respectively,
act on one of the orthogonal $m=1$ magnetic families according to the selection rules.
A similar diagram can be drawn for the orthogonal family.
All potential dynamo mechanisms can be represented by closed paths, or ``loops'', in this diagram.
Moreover, any dynamo loop must involve both poloidal and toroidal magnetic field.
It is then immediately clear from Fig.~\ref{fig:dynamo_mechanism} that the toroidal flow (dashed arrows)
alone cannot act as a dynamo, since then there would be no mechanism for the generation of 
poloidal magnetic field from toroidal magnetic field.
The simplest possible dynamo loops, 
i.e., those involving just two or three steps,
are drawn in Fig.~\ref{fig:dynamo_loops}.
We consider only the loops that contain the equatorial dipole
(in the set $\bp_\oo^\cc$) because the dynamo field is dominated by this 
component in all our simulations. 

\begin{figure}
 \centering
   \includegraphics[clip=true,width=10cm]{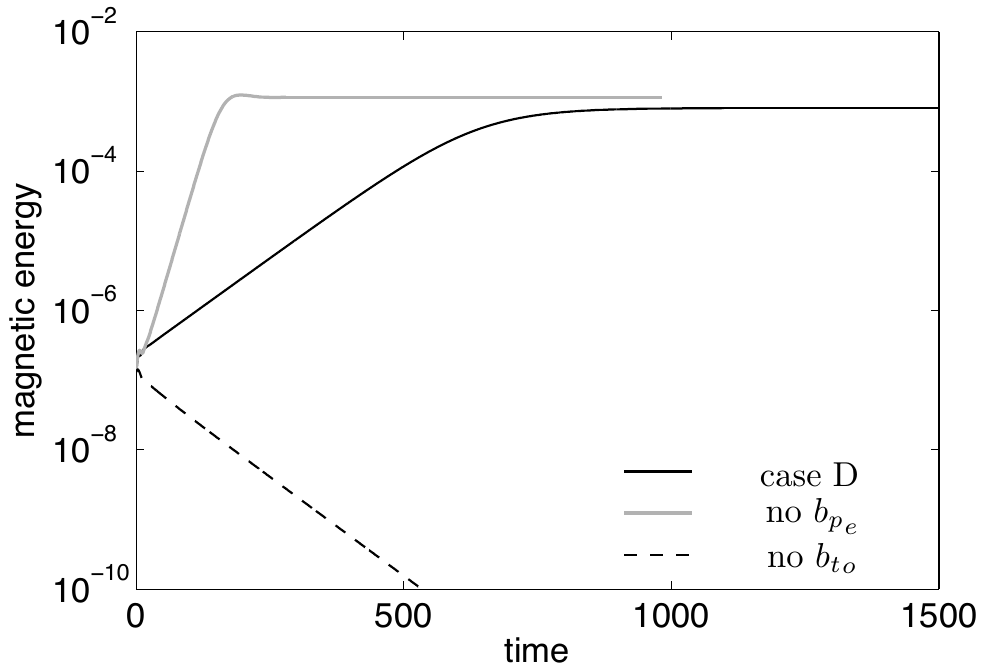}
  \caption{Times series of the magnetic energy in the cases where either the coefficients $\bp_\ee$ 
            or the coefficients $\bt_\oo$ of the $m=1$ magnetic field
	  are suppressed, compared with that for the full MHD simulation 
          in Case D. All the cases have similar wall parameters.
	  A global magnetic diffusion time is $\tau_{\eta}=\Rm=300$.}
  \label{fig:E_mag_test}
\end{figure}

To determine which of the dynamo loops in Fig.~\ref{fig:dynamo_loops}
most likely
represents the essential part of the dynamo mechanism,
we have performed a series of numerical experiments in which
all magnetic components in one of the quadrants shown in Fig.~\ref{fig:dynamo_mechanism} 
is artificially suppressed throughout the simulation.  That is, one of the sets
$\bp_\oo$, $\bp_\ee$, $\bt_\oo$, or $\bt_\ee$
is held at zero for all time (where the subscripts $\oo$ and $\ee$ imply all of the 
odd and even $l$ coefficients respectively but for the $m=1$ mode only).
Figure~\ref{fig:E_mag_test} shows time series of the magnetic energy
from Case D alongside corresponding time series
from the experiments where the coefficients $\bp_\ee$ and $\bt_\oo$ were suppressed.
When the spectral coefficients $\bp_\ee$ are suppressed, the flow still maintains
a dynamo, and the magnetic energy in the kinematic dynamo phase actually grows more 
rapidly than for the full MHD simulation (see further discussion in Section~\ref{sec:cond_wall}).
On the other hand, when the spectral coefficients $\bt_\oo$ are suppressed, the dynamo fails. 
These results, though not conclusive, suggest that $\bt_\oo$ 
is necessary for the dynamo mechanism, whereas $\bp_\ee$ is not.
Of the four loops shown in Fig.~\ref{fig:dynamo_loops}, only Loop B
is consistent with these observations.  This loop has three steps:
(1) the shearing of the equatorial dipole into toroidal field of odd degree by the 
differential rotation, $\ut_\ee$, 
(2) the twisting of toroidal field of odd degree into toroidal field of even degree by $\ut_\ee$, and 
(3) the regeneration of the equatorial dipole from the toroidal field of even degree by $\up_\ee$.
We emphasize that this loop description is an idealization of the dynamo mechanism,
because the components not contained in Loop B are nevertheless present in the full simulation, 
and must influence the dynamo process to some extent.
Moreover the poloidal flow and magnetic diffusion both act to rearrange the field within each quadrant, 
without changing the symmetry properties.
However, the loop we identify is consistent with the schematic of the dynamo mechanism proposed
by \citet{Nor06} (see their Fig.~4) for similar flows.  

We now examine the effects on the dynamo action resulting from changing the thickness, 
$\hat{h}$, electrical conductivity, $\sigma_\rr$, and magnetic permeability, $\mu_\rr$, of the wall.
We emphasize that we fix the magnetic Prandtl number of fluid at $\Pm_\ff=1$ and the boundary forcing at $\Rey=300$,
and so the magnetic Reynolds number of the fluid in each case is $\Rm=300$. 

\subsection{Effect of the wall conductivity and thickness}
\label{sec:mur1}

\subsubsection{Dynamo threshold}
\begin{figure}
 \centering
  \includegraphics[clip=true,height=6cm]{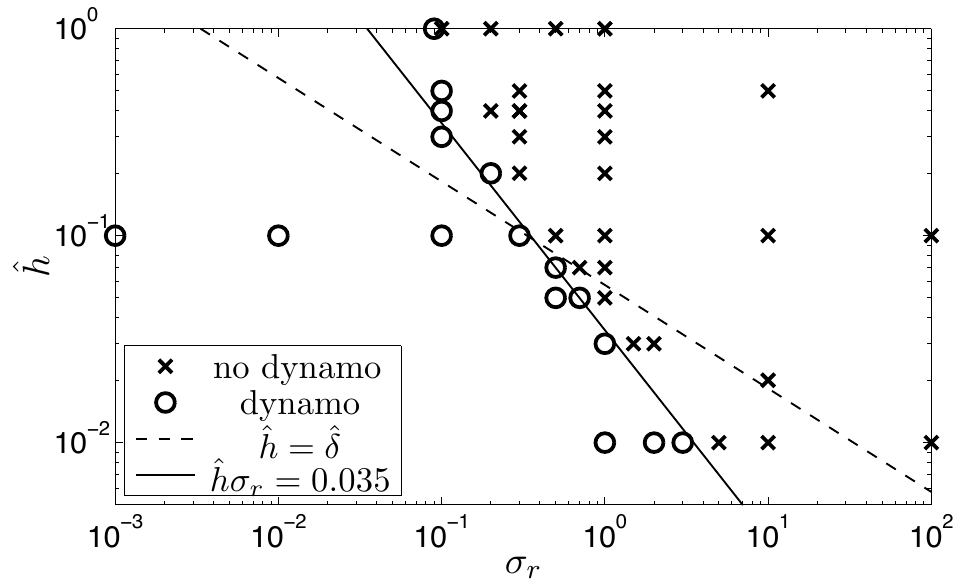}
  \caption{Results of dynamo simulations in the $(\sigma_\rr,\hat{h})$ space for $\mu_\rr=1$.
  The black line is $\hat{h} \sigma_\rr =0.035$, and
  the dashed line indicates where $\hat{h}$ is equal to the skin depth, $\hat{\delta} \propto \sigma_\rr^{-1/2}$
  (Eq.~(\ref{eq:skin})).}
  \label{fig:dynamo_space_mu1}
\end{figure}

\begin{figure}
 \centering
  \subfigure[]{\label{fig:ME_vs_time}
  \includegraphics[clip=true,width=6cm]{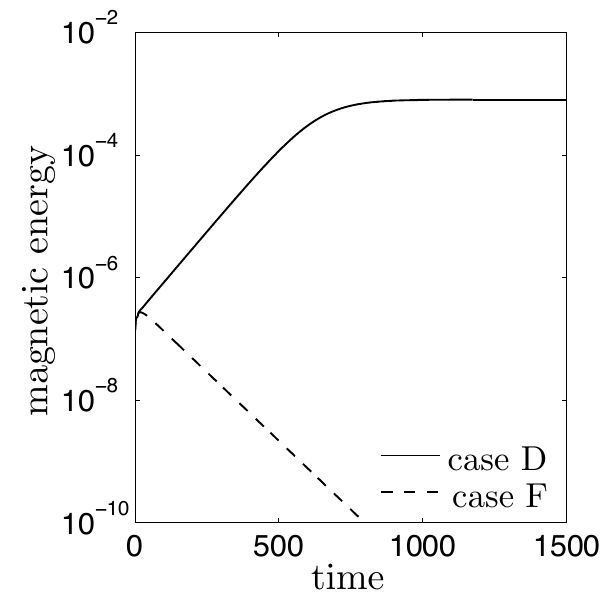}}
  \subfigure[]{\label{fig:Energy_vs_sig}
  \includegraphics[clip=true,width=7.5cm]{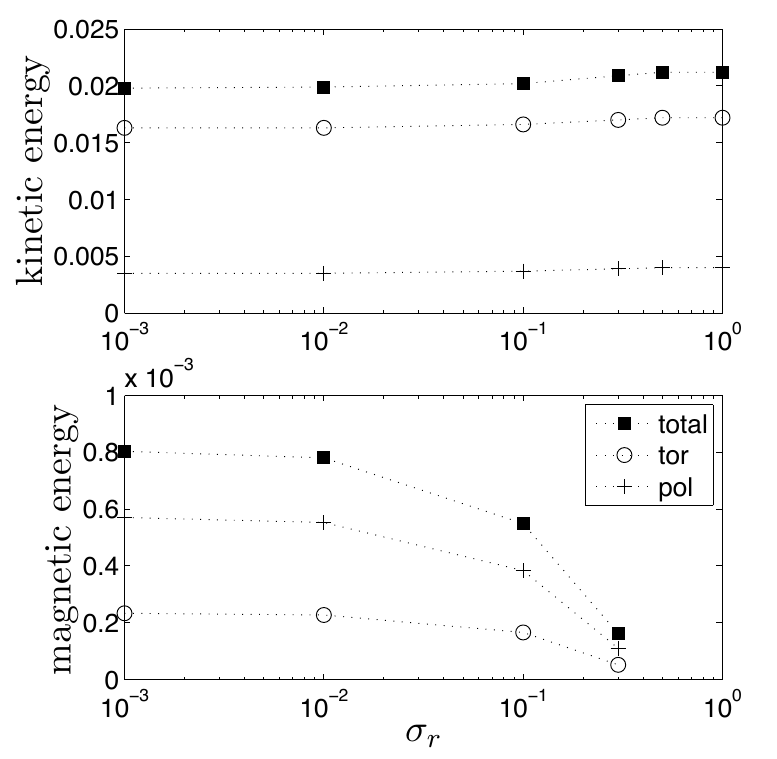}}
  \caption{
  (a) Times series of the magnetic energy in Case D $(\hat{h},\sigma_\rr)=(0.1, 10^{-3})$
  and in Case F $(\hat{h},\sigma_\rr)=(0.1, 1)$.
  (b) Steady saturated state kinetic energies (top) and magnetic energies (bottom) decomposed into total, toroidal and 
  poloidal parts for various $\sigma_\rr$ at $\hat{h}=0.1$, $\mu_\rr=1$. The values of the energies are averaged over the volume of the fluid.}
\end{figure}

First fixing the relative magnetic permeability at 
$\mu_\rr=1$, we have run
simulations for different values of $\sigma_\rr$ and $\hat{h}$,
varying both over several orders of magnitude.
Each simulation has been run for several global magnetic diffusion times,
which means several times the magnetic Reynolds number
$\Rm=300$ in non-dimensional units.
Figure~\ref{fig:dynamo_space_mu1} summarises where dynamo and non-dynamo states are found 
in the parameter space $(\sigma_\rr,\hat{h})$.  By ``dynamo state'' we mean that 
the magnetic energy in the corresponding simulation grows exponentially from the seed magnetic field 
used as initial condition, and then saturates at a significant steady value for the rest of the simulated 
time.
For non-dynamo cases, any initial field ultimately decays diffusively.
Figure~\ref{fig:ME_vs_time} shows the time series of the magnetic energy for two
cases representative of these behaviours: the dynamo case $(\hat{h},\sigma_\rr)=(0.1, 10^{-3})$ (Case D)
and a non-dynamo (failed) case $(\hat{h},\sigma_\rr)=(0.1,1)$ (hereafter called Case F).

It is clear from Fig.~\ref{fig:dynamo_space_mu1} that increasing either 
$\hat{h}$ or $\sigma_\rr$ is generally detrimental to dynamo action.  
For example, the cases with $\hat{h}=0.1$ and $\sigma_\rr \geq 0.5$ are all non-dynamos, whereas the cases with 
$\hat{h}=0.1$ and $\sigma_\rr \leq 0.3$ are all dynamos. This result is corroborated by Fig.~\ref{fig:Energy_vs_sig}, which plots
the values of the kinetic and magnetic energies in the fluid in the saturated phase for the simulations run at $\hat{h}=0.1$
for different $\sigma_\rr$.  The saturated magnetic energy decreases with increasing $\sigma_\rr$, and the dynamo 
disappears completely above $\sigma_\rr=0.3$.
In each dynamo simulation, about three quarters of the saturated magnetic energy comes from the poloidal field.
For $\sigma_\rr\leq 0.01$, the energy of the saturated magnetic field asymptotes to a limiting value, 
which is about 4\% of the total kinetic energy.
Although the dynamo cases have slightly lower kinetic energy than the
the non-dynamo cases, we find the flow structure to be almost identical
in all cases.

When the wall is sufficiently thin
($\hat{h}\lesssim0.1$ in Fig.~\ref{fig:dynamo_space_mu1})
the dynamo threshold closely follows the line $\hat{h}\sigma_\rr=0.035$.
For a thicker wall, the threshold in Fig.~\ref{fig:dynamo_space_mu1}
becomes significantly steeper, indicating that the dynamo mechanism is less sensitive to the wall thickness.
For $\hat{h}\gtrsim 1$ the dynamo threshold seems to asymptote to a limiting value of $\sigma_\rr\simeq 0.1$.

These results are consistent with those of \citet{Spe09}, who used the same values
of $\Rey$ and $\Pm_\ff$ but more idealized boundary conditions. 
They obtained a dynamo when the outer boundary was electrically insulating
($\sigma_\rr\to0$), 
but not when the outer boundary was perfectly conducting
($\sigma_\rr\to\infty$).
However, our results differ from those of \citet{Kha12}, 
who considered the asymptotic thin-wall limit of $\hat{h} \to 0$ with $\hat{h}\sigma_\rr$ finite.
An explanation for the discrepancy between 
their results and ours
is presented in Section~\ref{sec:analytical_sol}.

\subsubsection{Negative effect of the conducting wall}
\label{sec:cond_wall}

\begin{figure}
 \centering
   \subfigure[$\sigma_\rr=10^{-3}$, $\hat{h}=0.1$ (D)]{
    \raisebox{0.7cm}{\includegraphics[clip=true,width=4cm]{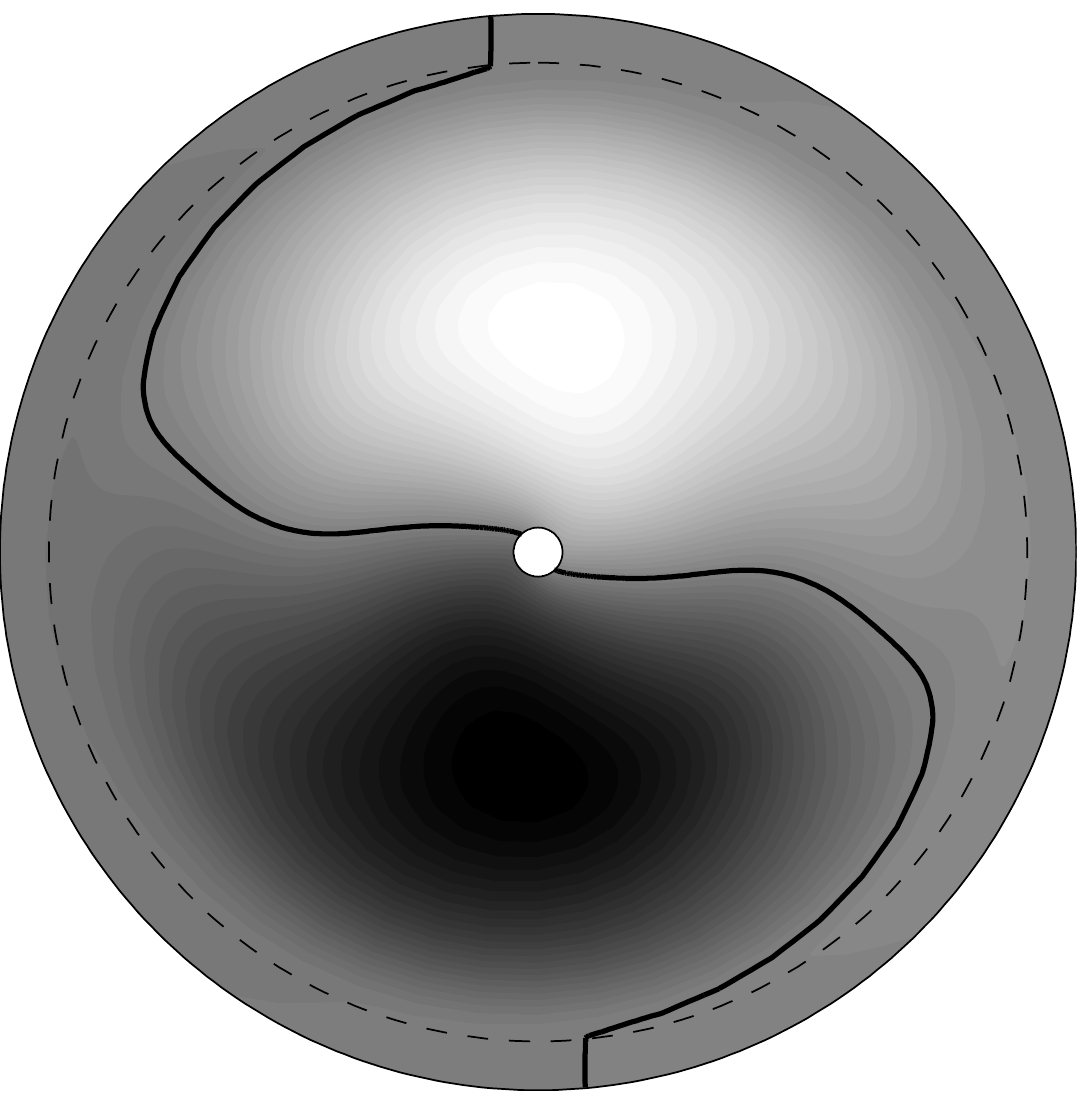}}\label{fig:Br_rphi_a}}
   \subfigure[$\sigma_\rr=1$, $\hat{h}=0.1$ (F)]{
   \raisebox{0.7cm}{\includegraphics[clip=true,width=4cm]{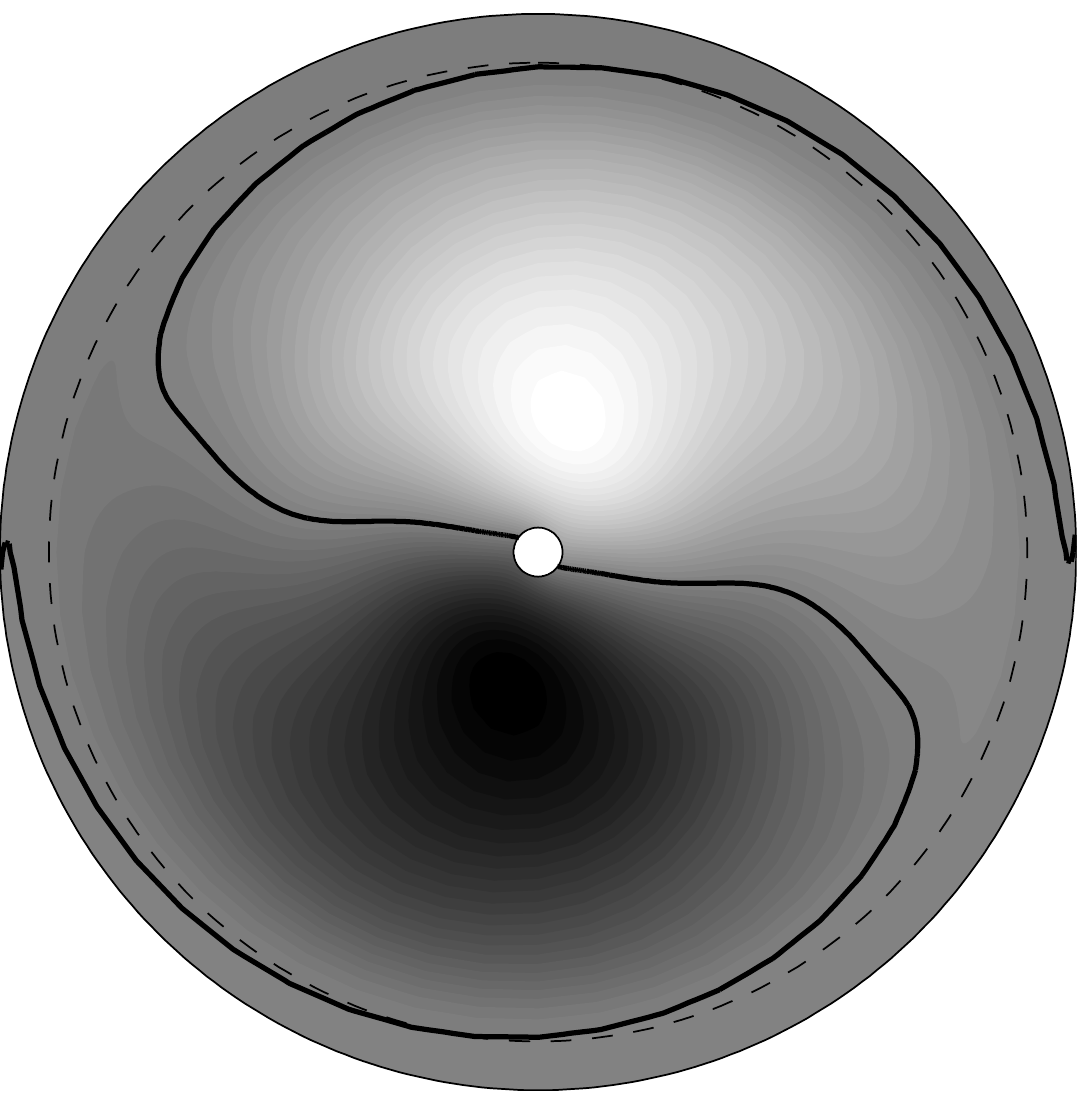}}\label{fig:Br_rphi_b}}
   \subfigure[$\sigma_\rr=1$, $\hat{h}=0.5$]{\label{fig:Br_rphi_c}
   \includegraphics[clip=true,width=5.4cm]{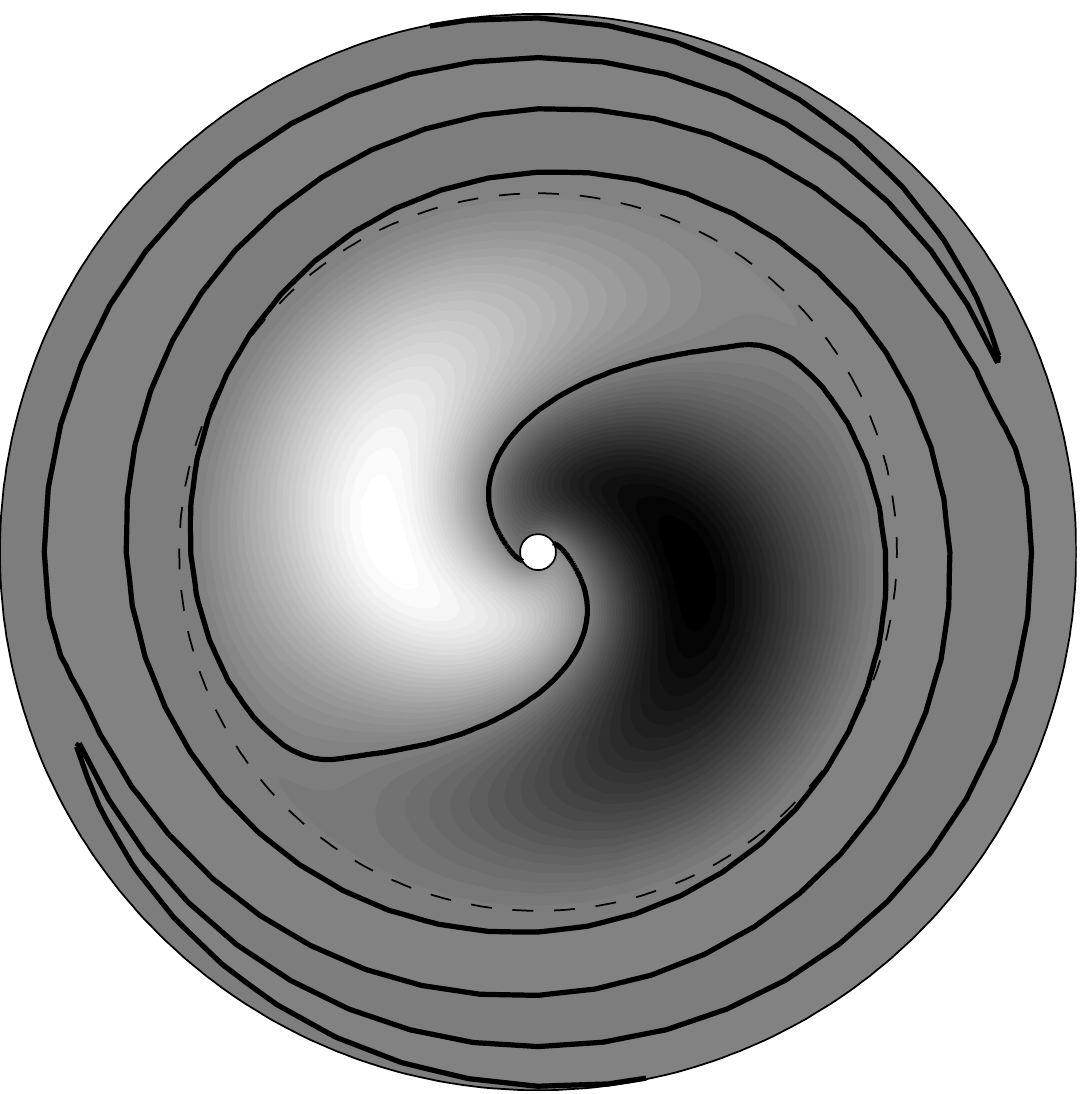}}
  \caption{Radial magnetic field $B_r$ in a conical section $(r,\phi)$ at colatitude $\theta=\pi/6$.
  The dashed line indicates the fluid--wall interface,
  and the solid bold line is the contour $B_r=0$
   (not a magnetic field line).}
  \label{fig:Br_rphi}
\end{figure}

The negative effect of a thick, highly conducting wall on dynamo action is explained by the 
induction of poloidal magnetic field in the wall.
For illustration,
Figure~\ref{fig:Br_rphi} shows
the radial component of the field, $B_r$, in a conical section $(r,\phi)$ at colatitude $\theta=\pi/6$
for the dynamo case, Case D, and for the failed dynamo case, Case F.
In Case F (Fig.~\ref{fig:Br_rphi_b}) the contours of $B_r$ in the wall spiral in the
direction of the wall rotation, as exhibited by the contour $B_r=0$,
which is plotted as a solid black line.  In Case D (Fig.~\ref{fig:Br_rphi_a}), by contrast, this contour is almost
exactly radial within the wall.
Note that the radial field $B_r$ is directly related to the poloidal scalar potential $\BP$ from Eq.~(\ref{eq:Br_Bp}).
Within the wall, the induction of poloidal field can be written
as an advection--diffusion equation for $B_r$
(see Eq.~(\ref{eq:Bpol_ss}) in Appendix~\ref{app:BC}).
For high wall conductivity $\sigma_\ww$, and hence low magnetic diffusivity,
the azimuthal advection of the poloidal field by the wall's differential rotation
becomes more significant, producing the spiral pattern seen in
Fig.~\ref{fig:Br_rphi_b}.
To show this pattern more clearly, we also plot in Fig.~\ref{fig:Br_rphi_c} a case with the same wall conductivity as Case F,
but with a thicker wall, $\hat{h}=0.5$.
In this simulation, the contour $B_r=0$ wraps several times around the sphere (Fig.~\ref{fig:Br_rphi_c}).
Because $\mu_\rr=1$, both $B_r$ and its radial derivative are continuous at the fluid--wall interface
(Eqs.~(\ref{eq:cont_bp}) and~(\ref{eq:cont_bs}))
and so the spiralling of the poloidal field in the conducting wall
is communicated directly to the fluid.

In terms of the Bullard--Gellman diagram in Fig.~\ref{fig:dynamo_mechanism},
the advection of the poloidal field by the toroidal
flow of the wall converts the equatorial dipole $\bp_\oo$ into
an equatorial quadrupole $\bp_\ee$,
and subsequently into an equatorial dipole of the opposite sign.
We note that advection of poloidal field by toroidal flow
is not part of the dynamo loop (Loop B in Fig.~\ref{fig:dynamo_loops})
responsible for maintaining the magnetic field.
In fact, this advection seems to be responsible for the failure of the dynamo in Case F.
The effect of the wall on the equatorial dipole can be interpreted physically
by noting that, as the wall conductivity is increased, the poloidal field
lines become increasingly ``anchored'' to the wall.
As a result, the counter rotation of the two hemispheres
``tears apart'' the magnetic field produced in the fluid, hindering 
the dynamo process.

This argument also provides a plausible explanation for
the behaviour seen in Fig.~\ref{fig:E_mag_test}:
for simulations with the same wall parameters as Case D,
the magnetic energy grows faster during the kinematic phase
if the spectral coefficients $\bp_\ee$ are suppressed.
Suppressing these coefficients prevents the advection of the poloidal field by the toroidal flow
(in the fluid and in the wall).

\begin{figure}
 \centering
 \subfigure[]{\label{fig:bp1}
   \includegraphics[clip=true,width=6cm]{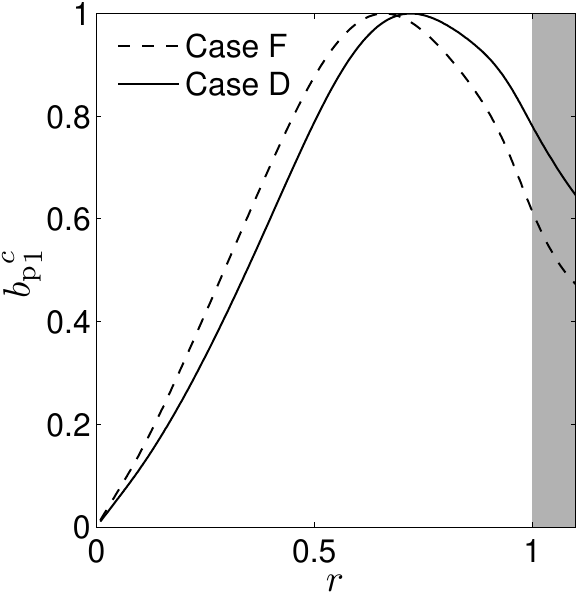}}
 \subfigure[]{\label{fig:jt1}
   \includegraphics[clip=true,width=6cm]{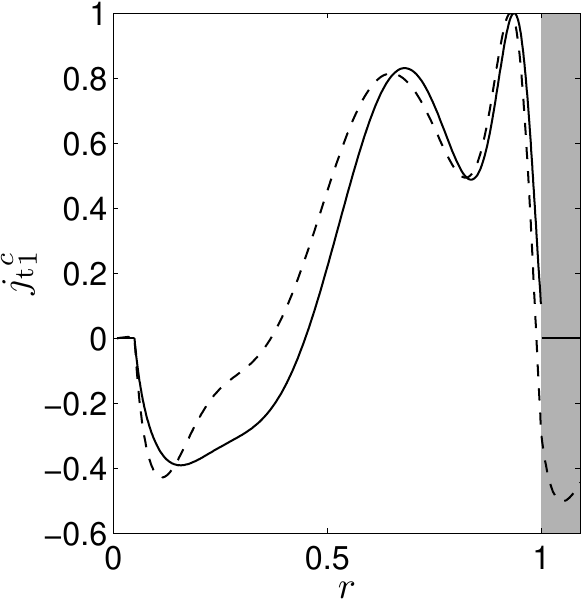}}
  \caption{Radial profiles of the spectral coefficients $\bp_1^{\cc}$ and $\jt_1^{\cc}$
  in Cases D and F.
  The profiles have been rescaled to have a maximum value of 1.
  The shading indicates the wall region.}
  \label{fig:r_profile_Bspectral}
\end{figure}

An alternative, but equivalent, physical interpretation concerns the circulation of electric currents
within the wall.
The relative motion of the wall and the (stationary) equatorial dipole
induces toroidal ``eddy'' currents in the wall, by a process analogous to
a skin effect.
Indeed, the degree of spiraling
in Figure~\ref{fig:Br_rphi}
can be measured in terms of the skin depth,
\begin{equation}
  \delta = (m\Omega(\theta)\sigma_\ww\mu_\ww)^{-1/2},
  \label{eq:skin}
\end{equation}
where $\Omega(\theta)$ is the angular velocity of the wall
and $m$ is the azimuthal order of the dominant magnetic mode (here $m=1$).
In dimensionless units, the minimum skin depth is
\begin{equation}
  \hat{\delta} = (\Rm\,\sigma_\rr\mu_\rr)^{-1/2}.
  \label{eq:skin2}
\end{equation}
This is approximately the radial separation between the two ``spiral arms''
of the contour $B_r = 0$.
The toroidal eddy
currents, in turn, induce opposing poloidal magnetic field within the fluid,
and the overall effect is to weaken the equatorial dipole. 
This is shown clearly in Fig.~\ref{fig:r_profile_Bspectral},
which plots radial profiles of the spectral coefficients
$\bp_1^{\cc}$ (corresponding to the equatorial dipole) and $\jt_1^{\cc}$ (corresponding to the toroidal electric current 
responsible for the induction of the equatorial dipole (Eq.~(\ref{eq:def_jt}))) for Cases D and F.
In Case F, the field decays exponentially with time,
and so the profiles have been rescaled to allow a direct comparison with Case D.
We find that the conducting wall in Case F allows the circulation of toroidal electric currents
that are reversed relative to the currents in the fluid,
thereby reducing the amplitude of the equatorial dipole
plotted in Fig.~\ref{fig:bp1}.

In Fig.~\ref{fig:dynamo_space_mu1}, the line \mbox{$\hat{h}=\hat{\delta}$} is plotted as the dashed line.
For \mbox{$\hat{h} < \hat{\delta}$}, the dynamo threshold is determined by the product $\hat{h}\sigma_\rr$
and not by the ratio $\hat{h}/\hat{\delta}$.
This means that the skin effect can be significant even if the wall thickness $\hat{h}$ is significantly 
smaller than the skin depth $\hat{\delta}$.
Physically, for $\hat{h}<\hat{\delta}$, the threshold depends on the amplitude of the opposing toroidal currents, which is proportional
to $\sigma_\rr$ (Eq.~(\ref{eq:cont_jt})) integrated over the wall thickness.
The line $\hat{h}=\hat{\delta}$ 
represents the boundary between the ``thin wall'' regime just described,
and the ``thick wall'' regime wherein the threshold becomes independent of $\hat{h}$.

\begin{figure}
 \centering
   \subfigure[]{\label{fig:bt1}
   \includegraphics[clip=true,width=6cm]{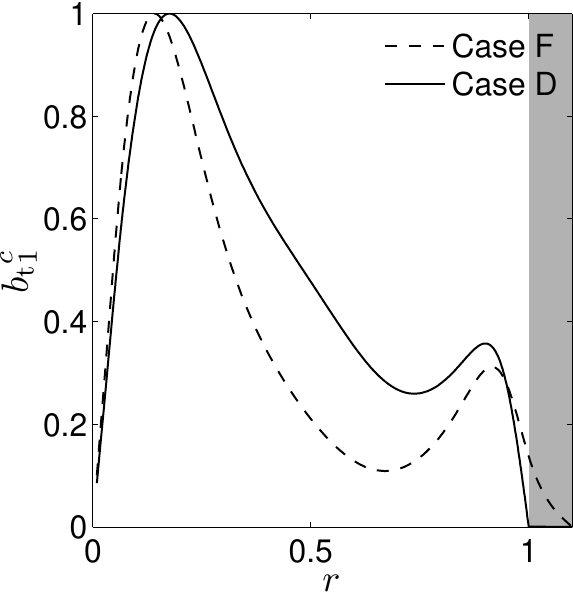}}
   \subfigure[]{\label{fig:bt2}
   \includegraphics[clip=true,width=6cm]{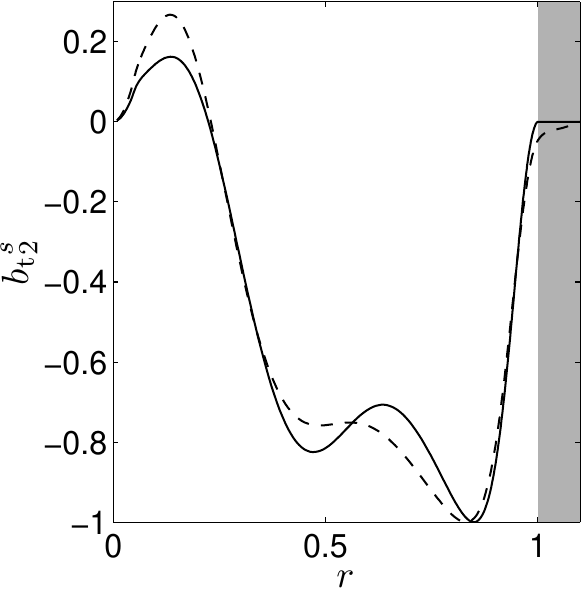}}
  \caption{Radial profiles of the spectral coefficients
  $\bt_1^{\cc}$ and $\bt_2^{\ss}$
  of the toroidal magnetic field for Cases D and F.
  The profiles have been rescaled to have a maximum absolute value of 1.}
  \label{fig:bt1bt2}
\end{figure}

Varying $\sigma_\rr$ and $\hat{h}$ also has consequences for the
poloidal currents, and hence for the toroidal magnetic field.
For an insulating wall ($\sigma_\rr=0$),
electric currents cannot flow out of the fluid,
and so we must have $\JP=0$ at $r=r_\oo$.
This implies, by Eq.~(\ref{eq:def_jp}),
that the toroidal magnetic field must also vanish at $r=r_\oo$,
and so is forced to a rapid decrease in the fluid region close to the wall.
Conversely, for a conducting wall, currents can leave the fluid
and recirculate within the wall, so 
the decrease of the toroidal field towards zero does not have to occur at $r=r_{\oo}$ 
but rather at the outer boundary with the vacuum.
In this way, the presence of a thick conducting wall promotes the generation
of toroidal field in the fluid, by allowing it to match to the vacuum boundary condition over a larger radial domain 
and so by shielding the fluid from the vacuum boundary condition.
To illustrate this, Figure~\ref{fig:bt1bt2} compares plots of 
the spectral coefficients of the toroidal field $\bt_1^{\cc}$ and $\bt_2^{\ss}$
from Cases D and F.
The decrease of $\bt_1^{\cc}$ and $\bt_2^{\ss}$ towards zero is indeed more rapid
in the outer part of the fluid for Case D.
However the generation of toroidal
field occurs mainly in the inner part of the domain, and so the influence of the
wall on the toroidal field is rather minor here.

In summary, we find that a thick conducting wall has competing effects 
on the generation of poloidal and toroidal magnetic fields.
A conducting wall allows a stronger toroidal magnetic field
to be generated in the outer part of the fluid, by permitting poloidal currents
to flow across the fluid--wall interface.
However, a conducting wall also allows the induction of toroidal eddy currents
in the wall that oppose the generation of non-axisymmetric poloidal magnetic field in the fluid.
In our simulations,
the negative effect of these eddy currents on the
poloidal field outweighs the positive effect on the toroidal field,
and so a conducting wall inhibits dynamo action.

\subsection{Effect of the wall permeability and thickness}
\label{sec:sigmar1}

\subsubsection{Dynamo threshold}

\begin{figure}
 \centering
  \includegraphics[clip=true,height=6cm]{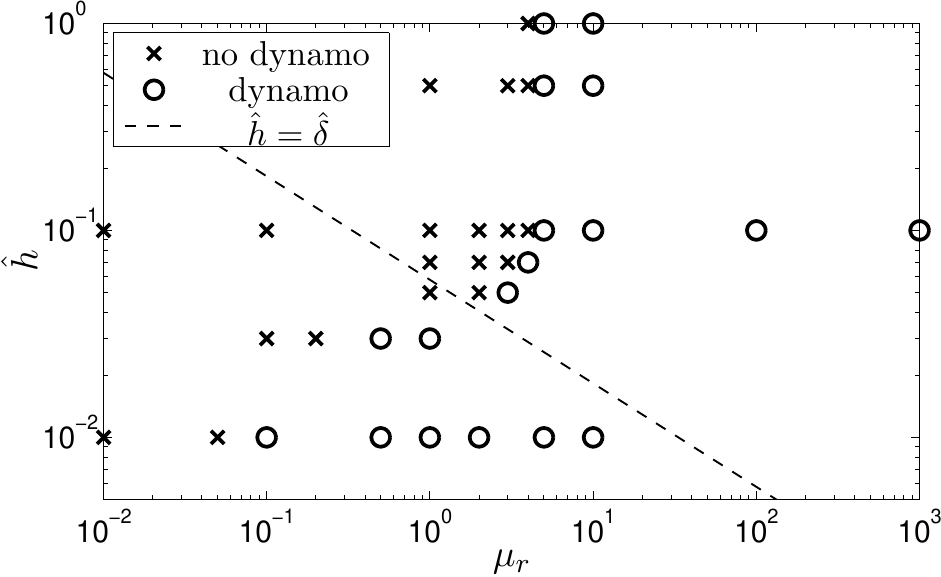}
  \caption{Dynamo simulations in the $(\mu_\rr,\hat{h})$ parameter space for $\sigma_\rr=1$.
  The dashed line indicates where $\hat{h}$ is equal to the skin depth, $\hat{\delta} \propto \mu_\rr^{-1/2}$.}
  \label{fig:dynamo_space_sig1}
\end{figure}

In this section, we fix
the relative conductivity of the wall at 
$\sigma_\rr=1$,
and present results from simulations with different values of $\mu_\rr$ and $\hat{h}$.
Figure~\ref{fig:dynamo_space_sig1} shows the location of the dynamo and non-dynamo simulations
in this parameter space.
In general, either increasing $\mu_\rr$ or 
decreasing $\hat{h}$ is favorable for dynamo action. 
For $\hat{h}\gtrsim0.1$, the dynamo threshold approaches a line with
$\mu_\rr\simeq5$, indicating that the dynamo mechanism becomes insensitive to the wall thickness.
For $\hat{h}<0.1$, the dynamo threshold does not follow an obvious power law.

\subsubsection{Positive effect of high magnetic permeability}

\begin{figure}
 \centering
   \includegraphics[clip=true,width=4cm]{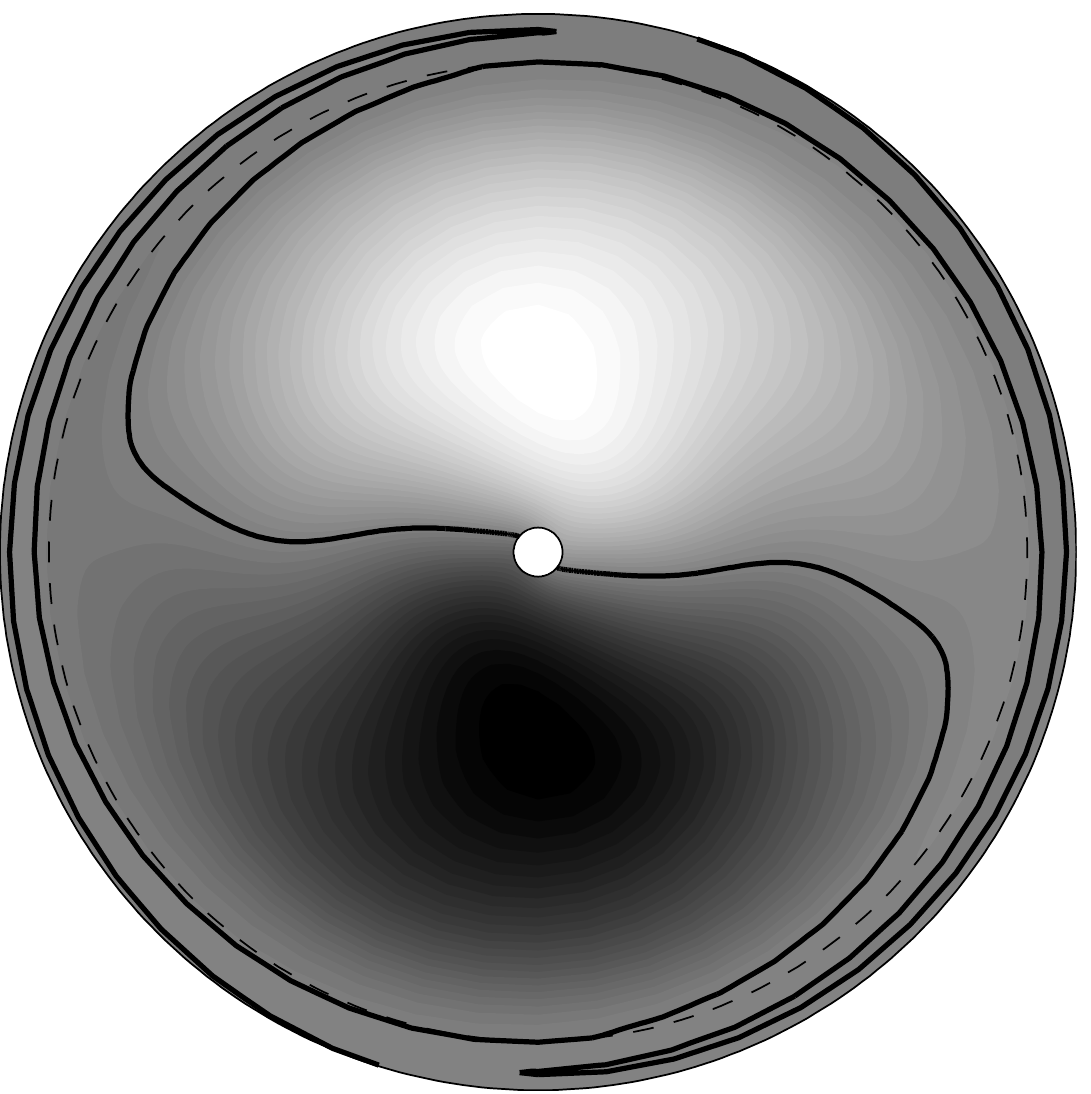}
  \caption{Same as Fig.~\ref{fig:Br_rphi} for $\mu_\rr=10$, $\sigma_\rr=1$ and $\hat{h}=0.1$.}
  \label{fig:Br_rphi_sig1}
\end{figure}

\begin{figure}
 \centering
 \subfigure[]{\label{fig:bp1_mu}
   \includegraphics[clip=true,height=6cm]{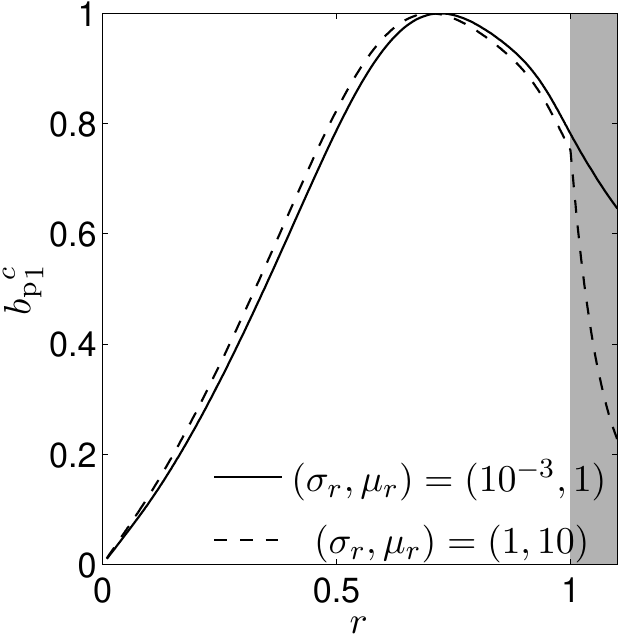}}
 \subfigure[]{\label{fig:bt1_mu}
   \includegraphics[clip=true,height=6cm]{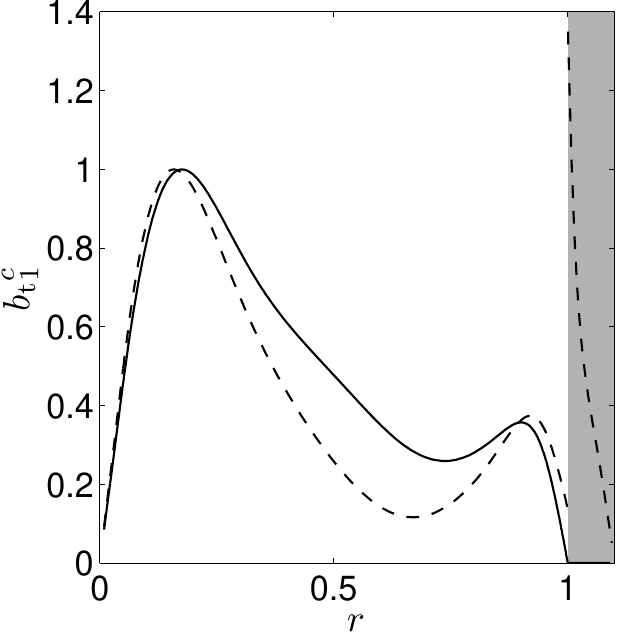}}
  \caption{Radial profiles of the spectral coefficients $\bp_1^\cc$ and $\bt_1^\cc$
  for the cases $(\sigma_\rr,\mu_\rr)=(10^{-3},1)$ 
  and $(\sigma_\rr,\mu_\rr)=(1,10)$.  Both cases have $\hat{h}=0.1$.
  The profiles are normalized against their maximum values in the fluid.}
\end{figure}

Figure~\ref{fig:Br_rphi_sig1} shows the radial component of the 
magnetic field, $B_r$, in a conical section $(r,\phi)$ 
at colatitude $\theta=\pi/6$
for the dynamo case $(\hat{h},\mu_\rr)=(0.1,10)$ in a similar manner to Fig.~\ref{fig:Br_rphi}.
The relatively high permeability, $\mu_\rr=10$,
implies a low magnetic diffusivity in the wall,
and leads to
spiraling of the radial magnetic field.
However, unlike the case of large $\sigma_\rr$, this spiraling in the wall
does not necessarily imply spiraling in the fluid,
because the radial derivative of $\BP$ is not continuous at $r=r_\oo$
(Eq.~(\ref{eq:cont_bs})).
Physically, this means that, whereas large $\sigma_\rr$ anchors the field lines to the wall,
large $\mu_\rr$ produces a paramagnetic ``suction'' of the tangential field components $B_\theta$
and $B_\phi$ into the wall \citep[e.g.][]{Gie12}.
In the limit $\mu_\rr\to\infty$, the matching condition (\ref{eq:Btan})
implies that the field in the fluid becomes perpendicular to the fluid--wall interface,
independently of any advection within the wall.
By decoupling the poloidal field in the fluid from that in the wall,
a dynamo field can be maintained in the fluid in spite of strong eddy currents in the wall.
Figure~\ref{fig:bp1_mu} compares the radial profiles of 
$\bp_1^\cc$ (corresponding to the equatorial dipole)
in the dynamo simulations with $(\hat{h},\sigma_\rr,\mu_\rr)=(0.1, 1, 10)$ and
$(\hat{h},\sigma_\rr,\mu_\rr)=(0.1,10^{-3},1)$ (Case D).
Although $\bp_1^\cc$ decays rapidly within the wall in the case with
$\mu_\rr=10$, 
the profiles in the fluid are very similar in both cases,
showing that the increase in $\mu_\rr$ compensates for the increase in $\sigma_\rr$.

The situation for the toroidal magnetic field is somewhat similar,
as shown in Fig.~\ref{fig:bt1_mu}.
Even though the high permeability of the wall allows for a much larger toroidal field in the wall,
the field in the fluid is not much affected.  In fact, the toroidal field in the fluid very closely matches
that seen in Case F (Fig.~\ref{fig:bt1}) after rescaling appropriately.

In summary,
a high magnetic permeability in the wall effectively decouples the field in the fluid from
that in the wall.  In our simulations, this decoupling promotes dynamo action, by preventing
the equatorial dipole from being torn apart by the differential rotation of the wall.
High wall permeability also allows a strong toroidal field to develop in the wall, but the toroidal
field in the fluid is not significantly affected.

\section{Analytical solution in a thin wall}
\label{sec:analytical_sol}

Our qualitative explanations for the effects of the wall properties on the dynamo can be
made rigorous in the asymptotic limit of vanishing wall thickness
considered by \citet{Rob10} and \citet{Kha12}.
Taking this limit also allows a more precise comparison of our results with those of \citeauthor{Kha12}.
The general form of the thin-wall boundary conditions is derived in Appendix~\ref{app:BC},
and compared with the special cases of \citeauthor{Rob10} and \citeauthor{Kha12}.

The derivation assumes that the wall thickness $h$ is much smaller than the characteristic 
scale of radial variations in the wall.
For the steady dynamo magnetic fields considered here,
and with a prescribed velocity in the wall of the form
\begin{equation}
  \vel =  r \sin \theta \, \Omega(\theta) \vect{e}_{\phi},
  \label{eq:uphi_wall}
\end{equation}
this assumption requires that $h$ is much smaller than the skin depth $\delta$ given by Eq.~(\ref{eq:skin}).
The boundary condition for each spherical harmonic coefficient of the poloidal magnetic scalar potential,
$\bp_l^m$, at $r=r_\oo^-$ is then
\begin{flalign}
  &&-\left. \frac{\dd\ln ( r \bp_l^m)}{\dd\ln r} \right|_{r_\oo^-} & = 
  \frac{l + \ii m \, \Rm \, \sigma_\rr \hat{h} A_l^m}{1+l\mu_\rr\hat{h}} ,&
  \label{eq:dBpdr}
  \\
  &\mbox{where}& A_l^m & =   \frac{\left[\hat{\Omega} \BP\right]_l^m}{\bp_l^m}.
\end{flalign}
Here,
$\hat{\Omega}(\theta) = \Omega(\theta) r_\oo/U_\ww$
is the dimensionless rotation rate,
and the notation $\left[\cdot\right]_l^m$ represents a  particular spherical harmonic component of degree $l$ and order $m$.

The first term in the numerator on the right-hand side
of Eq.~(\ref{eq:dBpdr}) arises from the vacuum boundary condition at $r=r_\oo+h$.
The second term in the numerator represents the contribution from the advection
of the poloidal magnetic field in the wall (or, equivalently, the induction of poloidal field by toroidal eddy 
currents in the wall).
Because $\hat{\Omega}$ is antisymmetric about the equator, $\hat{\Omega}\BP$ has the opposite equatorial symmetry to $\BP$.
The advection term therefore couples different degrees of the poloidal field, and, 
in particular, transforms the odd $l$ degrees  
into even $l$ degrees, 
which we showed was detrimental to dynamo action.
Since in all our simulations the poloidal magnetic field is dominated by the equatorial dipole component
$(l,m)=(1,1)$, we anticipate that the advection term will be significant in cases for which
$\sigma_\rr\hat{h} \gtrsim 1/\Rm \simeq 0.003$.
In fact, in Fig.~\ref{fig:dynamo_space_mu1} we find that
the dynamo threshold roughly follows the line $\sigma_\rr\hat{h} = 0.035$, that is $\sigma_\rr\hat{h} \simeq 10/\Rm$,
in the parameter space $\hat{h}<\hat{\delta}$ for which Eq.~(\ref{eq:dBpdr}) is valid.

Equation~(\ref{eq:dBpdr}) also demonstrates how a large value of $\mu_\rr$ can offset
the negative effect of a large value of $\sigma_\rr$.
Indeed, in the limit $\mu_\rr\hat{h} \to \infty$ the right-hand side of Eq.~(\ref{eq:dBpdr})
vanishes, implying that the poloidal field lines become perpendicular to the
fluid--wall interface, even if the advection term dominates the numerator.
Figure~\ref{fig:dynamo_space_sig1} shows that, for $\sigma_\rr=1$, a permeability of $\mu_\rr=5$ is enough
to maintain a dynamo when $\hat{h}\ge0.1$.
However, note that Eq.~(\ref{eq:dBpdr}) is not strictly valid in this regime, since the wall thickness
is larger than the skin depth.

As discussed in Appendix~\ref{app:BC},
the boundary condition~(\ref{eq:dBpdr}) differs from the boundary condition for the poloidal magnetic field
used in \citet{Kha12} in two respects.
Firstly, in their model the outer wall is at rest, and so the advection term is absent.
Secondly, they considered only the kinematic phase in which the magnetic field grows or decays
exponentially, and so their boundary condition contains an additional $\partial\bp/\partial t$ term.
The absence of the advection term means that there is no skin effect in their model, and so the conductivity
of the wall has little effect on the dynamo process in their model.

The boundary condition for the spherical harmonic coefficients of the toroidal magnetic scalar potential, 
$\bt^m_l$, at $r=r_\oo^-$ is  
\begin{eqnarray}
	 -\left. \frac{\partial \ln( r \bt^m_l)}{\partial \ln r} \right|_{r_\oo^-} & = &
	 \frac{1}{\sigma_\rr \hat{h}} ,
	\label{eq:dBtdr}
\end{eqnarray}
which is identical to that of \citet{Rob10}.
This explains the insensitivity of the toroidal field to the permeability of the wall in our results.
For small values of $\sigma_\rr \hat{h}$
we recover the insulating boundary condition $\BT=0$,
implying no radial currents at the fluid--wall interface.
For finite values of $\sigma_\rr \hat{h}$, a finite
radial current is permitted to the extent that the current can recirculate within the wall.
Equation~(\ref{eq:dBtdr}) imposes that the radial component of the current is proportional to the divergence
of the angular components, with a constant of proportionality given by the radially integrated conductivity.

\section{Conclusions and discussion}

We have performed a series of numerical simulations to study dynamo action
generated by a steady, hydrodynamically-stable, laminar axisymmetric shear flow
driven by the counter-rotating hemispheres of a spherical shell.  
We have studied the effects of varying independently the thickness, $\hat{h}$, 
electrical conductivity, $\sigma_\rr$, and 
magnetic permeability, $\mu_\rr$, of the outer wall on the dynamo action.
For certain favorable magnetic boundary conditions, the flow maintains a magnetic 
field consisting mainly of a stationary equatorial dipole and a toroidal
component, both of which have azimuthal symmetry $m=1$.

The effects on the dynamo action of changing independently the parameters of the outer wall 
are summarized in Table~\ref{tab:summary}. The table emphasizes the effect that each change 
has on the main poloidal and toroidal components.

\begin{table}
\centering
\begin{tabular}{c c c c c}
\hline
\hline
& \multicolumn{2}{c}{{Laminar flow ($\Rey=300$)}}  & \multicolumn{2}{c}{{Turbulent flow ($\Rey=50000$)}}
\\ \hline
& poloidal field & toroidal field &  poloidal field & toroidal field
\\
& $m=1$ & $m=1$ & $m=0$ & $m=0$
\\ \hline
$\sigma_\rr \nearrow$ & $-$  & $+$ & & $+$ 
\\
 & eddy currents & buffer from vacuum &  & buffer from vacuum
\\ \hline
$\mu_\rr \nearrow$ & $+$ & & & $+$
\\
 &  field normal at $r=r_\oo^-$ & & & enhance $\omega$-effect
\\ \hline
$\hat{h} \nearrow$ & $-$ if $\hat{h}<\hat{\delta}$& $+$ & & $+$ 
\\
 & eddy currents & buffer from vacuum & & buffer from vacuum
\\ \hline \hline
\end{tabular}
\caption{Summary of the effect of the parameters of the outer wall on the 
generation of the poloidal and toroidal magnetic fields
in the case of the laminar flow considered in this paper, and 
compared to the results obtained with a similar boundary-driven flow at larger Reynolds number in \citet{Gue12}.
The $+$ and $-$ symbols indicate whether the effect is favorable or detrimental to the generation of the field.}
\label{tab:summary}
\end{table}

In general, we
 find that decreasing the wall thickness, decreasing the wall conductivity and increasing the wall 
permeability all promote dynamo action in this system.  
For high wall conductivity or permeability, 
the advection of the poloidal field by the rotation of the wall can be described 
as the induction of toroidal ``eddy'' currents in the wall by the poloidal field.
For large $\sigma_\rr$, 
the eddy currents oppose the poloidal field in the fluid, in a manner analogous to a skin effect,
and are detrimental to dynamo action.
The skin depth, $\hat{\delta}$, which is proportional to 
\mbox{$(\sigma_\rr \mu_\rr)^{-1/2}$}, determines the transition between the thick wall regime, 
$\hat{h}>\hat{\delta}$,  and the thin wall regime, $\hat{h}<\hat{\delta}$.
In the thick wall regime, the dynamo threshold becomes independent of $\hat{h}$. 
In the thin wall regime with homogeneous permeability, 
the dynamo threshold depends on the radially-integrated conductivity $\hat{h}\sigma_\rr$.
Increasing the wall magnetic permeability is favorable to dynamo action because
high wall permeability constrains the magnetic field in the fluid to be normal to the interface 
with the wall. This effectively disconnects the fluid from any eddy currents in the wall.

Increasing either the conductivity or the thickness of the wall allows stronger tangential currents in the wall and, 
by continuity, stronger values of the radial current and toroidal field at the fluid-wall interface.  
A highly conducting (or thick) wall thus creates a buffer region between the fluid and the vacuum outside.  
However, allowing stronger values of the toroidal field at the fluid--wall interface is favorable to dynamo action only
if the velocity shear layer, where toroidal field is produced, is located close to the wall. 
For the laminar flows studied here, the shear layer extends across most of the bulk of the fluid interior,
and the positive effect of a thick conducting wall on the toroidal field generation is
outweighed by the negative effect on the poloidal field generation.

It is interesting to compare our results with those of \citet{Kai99}, who studied the dynamo action of 
a helical flow surrounded by a (stationary) conducting wall.  They observed
the existence of an optimal thickness that minimizes the critical magnetic Reynolds number.
This is because, in their case, the positive effect of 
the penetration of radial currents into the wall (that is, of the buffer region for the toroidal field) 
outweighs the negative effect of the eddy currents as long as the wall is thinner than this optimal thickness.
This is an important difference from our study, where we find that
for walls significantly thinner than the skin depth, the skin effect still hinders the dynamo action.
This difference likely arises because, in the \citeauthor{Kai99} model, there is a shear discontinuity between 
the fluid and the wall, and so a conducting wall significantly enhances the generation of toroidal field.

Interestingly, the dependence of the dynamo threshold on the wall parameters for the 
laminar axisymmetric flows considered here differs from that found for 
turbulent flows at higher Reynolds number but with a similar azimuthal boundary forcing.
As summarized in Table~\ref{tab:summary}, in the turbulent case, increasing $\sigma_\rr$, $\mu_\rr$, 
or $\hat{h}$ is favorable for dynamo action \citep{Rob10,Gue12}.  
However, these turbulent dynamos have a distinctly different geometry: they are predominantly 
steady and axisymmetric ($m=0$), whereas the laminar dynamos considered here are necessarily non-axisymmetric.
This difference is significant, because an axisymmetric steady field is not subject to a skin effect.  
Indeed, \citet{Gue12} showed that the only significant effect of the outer wall on the 
dynamo action in the turbulent case is to support the generation of a strong axisymmetric toroidal field, 
which then feeds the other components of the dynamo cycle.
For large Reynolds numbers, the velocity shear layer created by the boundary forcing 
is narrow and confined close to the wall,
and the buffering effect of a highly conductive or thick wall allows this boundary layer 
to create toroidal field very efficiently by the $\omega$-effect.
A high wall permeability also enhances the $\omega$-effect in the turbulent case, 
by promoting a radial field at the fluid--wall interface adjacent to the shear layer through 
the same paramagnetic suction seen in the laminar case.

Ultimately then, the effect of magnetic boundary conditions in a dynamo model
depends on the geometry of the magnetic field as much as on the physical configuration of  
the model.  
In particular, a highly conducting, differentially rotating boundary tends to promote axisymmetric steady field 
configurations, and inhibit non-axisymmetric configurations.
Although this study was motivated by upcoming dynamo experiments, the understanding established is relevant
also to astrophysical dynamos.
For example, the presence of a differentially rotating conducting layer in Saturn has previously been 
invoked as an explanation for its highly axisymmetric magnetic field \citep{Ste82b,Sta10b}.
In this scenario,
the conducting ``wall'' is a stably stratified layer of fluid surrounding the deeper 
convective region in which the dynamo operates.  The differential rotation in this layer is produced 
by thermal winds arising from the latitudinal temperature gradient at the planet's surface.
\citet{Sta10b} found that the role of the stable layer on the axisymmetry of the magnetic field
depends on the equatorial symmetry of the thermal winds. This result can be explained using
our thin-wall boundary condition for the poloidal magnetic field (Eq.~(\ref{eq:dBpdr})).

\section*{Acknowledgments}
The authors would like to thank P.~H.~Diamond, C.~Forest, G.~A.~Glatzmaier, G.~R.~Tynan
for useful discussions and two anonymous referees for improving the manuscript.
Financial support was provided by the Center for Momentum Transport and Flow Organization (CMTFO), 
a Plasma Science Center sponsored by the US Department of Energy (DoE) Office of Fusion Energy Sciences 
and the American Recovery and Reinvestment Act (ARRA) 2009.
This research was further supported by an allocation of advanced computing resources provided by the 
National Science Foundation (NSF).  The computations were performed on the NSF Teragrid/XSEDE machine Kraken 
at the National Institute for Computational Sciences (NICS).

\appendix

\section{The magnetic boundary conditions in the thin-wall limit}
\label{app:BC}
We consider a spherical wall of thickness $h$, electrical conductivity $\sigma_\ww$, 
and magnetic permeability $\mu_\ww$, separating a fluid with $\sigma=\sigma_\ff$ and $\mu=\mu_\ff$ 
from an external vacuum with $\sigma=0$ and $\mu=\mu_0$. 
In the limit $h\to0$ we anticipate that the effect of the wall depends only on the radially 
integrated conductivity $h\sigma_\ww$ and permeability $h\mu_\ww$ \citep{Rob10}.

We suppose that the wall is differentially rotating with angular velocity $\Omega(\theta)$.
Within the wall, the magnetic induction equation (\ref{eq:induction}) then takes the form
\begin{equation}
  \frac{\partial\B}{\partial t} = \vn\times(\Omega r\sin\theta\,\vect{e}_\phi\times\B - \eta_\ww\nabla\times\B)
  \label{eq:ind_wall}
\end{equation}
where $\eta_\ww = 1/(\sigma_\ww\mu_\ww)$.

\subsection{Poloidal magnetic field}
The radial component of Eq.~(\ref{eq:ind_wall}) can be written as
an advection--diffusion equation for $B_r$:
\begin{align}
  \left(
    \frac{\partial}{\partial t}+\Omega\frac{\partial}{\partial\phi}
  \right)(r^2B_r) &= \eta_\ww\left[\frac{\partial^2}{\partial r^2} - \frac{1}{r^2}L_2\right](r^2B_r),
	\label{eq:Bpol_ss}
\end{align}
where $L_2$ is the angular Laplacian operator defined by Eq.~(\ref{eq:L2}).
For convenience we introduce a new variable $P = r^2B_r$;
the matching conditions~(\ref{eq:cont_bp}) and (\ref{eq:cont_bs})
then imply that $P$ and $\mu^{-1}\partial P/\partial r$ are
continuous.

Since Equation~(\ref{eq:Bpol_ss}) has no explicit dependence on
either $t$ or $\phi$, it is convenient to decompose $P$ spectrally
in those coordinates and then solve for each mode separately.
We therefore assume that $P \propto \ee^{\ii\omega t + \ii m\phi}$,
for some constants $\omega$ and $m$.
If we also assume that the thickness of the wall is much smaller than the scale of any latitudinal or azimuthal variations within the wall, 
then Eq.~(\ref{eq:Bpol_ss}) can be approximated as
\begin{equation}
  \ii\frac{P}{\delta^2} \simeq
  \frac{\partial^2P}{\partial r^2},
  \label{eq:Bpol_ss_approx}
\end{equation}
where $\delta(\theta)$ is a generalized skin depth
\begin{equation}
    \delta(\theta) = \pleft \frac{\eta_\ww}{\omega + m \Omega(\theta)} \pright^{1/2}.
\end{equation}
We emphasize that $\delta$ depends on colatitude, as well as on the frequency $\omega$ and azimuthal wavenumber $m$.  
We are interested here in the regime with $h \ll \delta$, in which case we can
approximate the radial dependence of $P$ within the wall via Taylor expansion.  In particular, we have
\begin{align}
  \left.P\right|_{r=(r_\oo+h)^-} \;\; &= \;\; \left.P\right|_{r=r_\oo^+}
    + \left.\frac{\partial P}{\partial r}\right|_{r=r_\oo^+}h
    \;\; + \;\; O(h^2/\delta^2) ,
    \label{eq:Taylor1} \\
  \left.\frac{\partial P}{\partial r}\right|_{r=(r_\oo+h)^-} \;\; &= \;\; \left.\frac{\partial P}{\partial r}\right|_{r=r_\oo^+}
    + \left.\frac{\partial^2P}{\partial r^2}\right|_{r=r_\oo^+}h
    \;\; + \;\; O(h^2/\delta^2),
  \label{eq:Taylor2}
\end{align}
where the superscripts $-$ and $+$ indicate points immediately inside and
outside a given radius respectively.
The second derivative of $P$ in Eq.~(\ref{eq:Taylor2}) can be inferred from Eq.~(\ref{eq:Bpol_ss_approx}).
We can then use the fact that $P$ and $\mu^{-1}\partial P/\partial r$ are continuous at $r=r_\oo^-$ and $r=(r_\oo+h)^+$ 
(Eqs.~(\ref{eq:cont_bp}) and~(\ref{eq:cont_bs})) to relate values just outside the wall in the vacuum and in the fluid: 
\begin{align}
  \left.P\right|_{r=(r_\oo+h)^+} &\simeq \left.P\right|_{r=r_\oo^-}
    + \frac{\mu_\ww}{\mu_\ff}\left.\frac{\partial P}{\partial r}\right|_{r=r_\oo^-}h, \label{eq:BC1_pol} \\
  \frac{\mu_\ww}{\mu_0}\left.\frac{\partial P}{\partial r}\right|_{r=(r_\oo+h)^+} &\simeq \frac{\mu_\ww}{\mu_\ff}\left.\frac{\partial P}{\partial r}\right|_{r=r_\oo^-}
    + \left.\frac{\ii}{\delta^2}P\right|_{r=r_\oo^-}h.
    \label{eq:BC_dpol}
\end{align}
Finally, we use the fact that, within the vacuum, we have
\begin{equation}
  \left[\frac{\partial P}{\partial r}\right]_l^m =
  -\frac{l}{r}\left[P\right]_l^m, \label{eq:vacuum}
\end{equation}
where the notation $\left[\cdot\right]_l^m$ represents a particular spherical harmonic component.  
From Eqs.~(\ref{eq:BC1_pol})--(\ref{eq:vacuum}) and the definition of $\delta$ 
we deduce the following boundary condition for the fluid at $r=r_\oo^-$:
\begin{equation}
  -\frac{l}{r_\oo}\frac{\mu_\ww}{\mu_0}\pleft\left[P\right]_l^m
    + \frac{\mu_\ww}{\mu_\ff}\left[\frac{\partial P}{\partial r}\right]_l^mh\pright
  \simeq \frac{\mu_\ww}{\mu_\ff}\left[\frac{\partial P}{\partial r}\right]_l^m
    + \left[\pleft\frac{\ii\omega + \ii m\Omega}{\eta_\ww}\pright P\right]_l^mh.
  \label{eq:BC_pol_full}
\end{equation}
If the wall is at rest ($\Omega(\theta)=0$) then we recover the thin-wall
boundary condition of \citet{Kha12}.
In general, it is easiest to implement Eq.~(\ref{eq:BC_pol_full})
as a dynamic boundary condition, by replacing $\ii\omega P$ by
$\partial P/\partial t$.  We then have
\begin{equation}
  \frac{\partial}{\partial t}\left[P\right]_l^m + \ii m\left[\Omega P\right]_l^m
  \simeq
  -\frac{l}{r_\oo}\frac{1}{\mu_0\sigma_\ww h}\left[P\right]_l^m
    -\frac{1}{\mu_\ff\sigma_\ww h}\pleft 1 +\frac{l}{r_\oo}\frac{\mu_\ww h}{\mu_0}\pright\frac{\partial}{\partial r}\left[P\right]_l^m.
  \label{eq:BC_dynamic}
\end{equation}
Since the coefficients on the right-hand side of (\ref{eq:BC_dynamic})
are both negative, this boundary condition is well posed.

If the rotation rate of the wall matches that of the fluid
next to the wall, then Eq.~(\ref{eq:Bpol_ss}) also applies
at the surface $r=r_\oo^-$, except with $\eta_\ww$ replaced by
$\eta_\ff = 1/(\sigma_\ff\mu_\ff)$.
In that case, we can rewrite the left-hand side of (\ref{eq:BC_dynamic}) as
follows:
\begin{equation}
  \eta_\ff\left[\frac{\partial^2P}{\partial r^2} - \frac{l(l+1)}{r_0^2} P\right]_l^m
  \simeq
  -\frac{l}{r_\oo}\frac{1}{\mu_0\sigma_\ww h}\left[P\right]_l^m
    -\frac{1}{\mu_\ff\sigma_\ww h}\pleft 1 +\frac{l}{r_\oo}\frac{\mu_\ww h}{\mu_0}\pright\left[\frac{\partial P}{\partial r}\right]_l^m.
  \label{eq:BC_Wentzell}
\end{equation}
This almost exactly matches the thin-wall boundary condition derived by
\citet{Rob10}.  However, their boundary condition includes spurious
terms proportional to $h^2$, which arise because they include $O(h^2/\delta^2)$ terms in the 
Taylor series (\ref{eq:Taylor1}), but not in (\ref{eq:Taylor2}).
Including such higher order terms in both (\ref{eq:Taylor1}) and (\ref{eq:Taylor2}) 
increases the complexity of the derivation,
but the results presented here are still obtained in the thin-wall limit
$h \to 0$ with $\sigma_\ww h$ and $\mu_\ww h$ both finite.

For the steady-state dynamos considered in this paper,
with $\mu_\ff = \mu_\ww/\mu_\rr = \mu_0$,
boundary condition (\ref{eq:BC_dynamic}) becomes
\begin{equation}
  -\left. \frac{\partial\ln\left[P\right]_l^m}{\partial\ln r} \right|_{r=r_\oo^-}
    \simeq \dfrac{l + \ii m\,\Rm\,\sigma_\rr\hat{h}\dfrac{[\hat{\Omega}P]_l^m}{\left[P\right]_l^m}}{1+l\mu_\rr\hat{h}},
\end{equation}
where $\hat{h}$ and $\hat{\Omega}$ are the thickness and 
angular velocity of the wall in non-dimensional units and $\Rm=U_\ww r_\oo \mu_0 \sigma_\ff$.

\subsection{Toroidal magnetic field}
A thin-wall boundary condition for the toroidal field can
be derived by taking the radial component of the curl of Eq.~(\ref{eq:ind_wall}),
and defining $T=r^2J_r$.
The derivation then
follows the same lines as in the previous section, with two modifications:
\begin{itemize}
  \item we replace $P \to T$ and $\mu \to \sigma$, except in the vacuum, where $\sigma=0$;
  \item the toroidal equivalent of Eq.~(\ref{eq:Bpol_ss}) has additional terms involving the shearing of field lines,
   which introduces additional terms into Eq.~(\ref{eq:BC_dpol}).
\end{itemize}
However, because the vacuum has $\sigma=0$, $T$ must vanish at $r = r_\oo+h$, and so the toroidal equivalent of 
Eq.~(\ref{eq:BC1_pol}) is simply
\begin{align}
  0 \simeq \left.T\right|_{r=r_\oo^-}
    + \frac{\sigma_\ww}{\sigma_\ff}\left.\frac{\partial T}{\partial r}\right|_{r=r_\oo^-}h.
\end{align}
From this we deduce immediately that the boundary condition for the toroidal field is
\begin{equation}
  \frac{\partial\ln\left[T\right]_l^m}{\partial\ln r}
    \simeq -\frac{1}{\sigma_\rr\hat{h}}.
\end{equation}
This exactly matches the thin-wall boundary condition of \citet{Rob10}.

\end{document}